\documentclass{aa}

\usepackage{natbib}
\bibpunct{(}{)}{;}{a}{}{,} % to follow the A&A style

\usepackage{graphicx}
\usepackage{txfonts}
\usepackage{hyperref}

\usepackage{lipsum}

\graphicspath{{images/}}

\usepackage{xcolor}

\defcitealias{bayeslosvd}{FBM21}
\defcitealias{Hinterer23}{H23}

\begin{document}

%
% Abbreviations used in the main .tex-file.
%

% for integrals
\newcommand{\dd}{\,\mathrm{d}}

% Useful things
\newcommand{\norm}[1]{\left\|#1\right\|}
\newcommand{\abs}[1]{\left\vert#1\right\vert}
\newcommand{\spr}[1]{\left\langle\,#1\,\right\rangle}
\newcommand{\kl}[1]{\left(#1\right)}
\newcommand{\Kl}[1]{\left\{#1\right\}}
\newcommand{\vl}{\, \vert \,}

\newcommand{\xv}{\boldsymbol{x}}
\newcommand{\yb}{\bar{y}}

\newcommand{\vmin}{{v_\text{min}}}
\newcommand{\vmax}{{v_\text{max}}}
\newcommand{\zmin}{{z_\text{min}}}
\newcommand{\zmax}{{z_\text{max}}}
\newcommand{\tmin}{{t_\text{min}}}
\newcommand{\tmax}{{t_\text{max}}}
\newcommand{\lmin}{{\lambda_\text{min}}}
\newcommand{\lmax}{{\lambda_\text{max}}}
\newcommand{\umin}{{u_\text{min}}}
\newcommand{\umax}{{u_\text{max}}}

\newcommand{\expect}{\mathbb{E}}
\newcommand{\stddev}{\mathbb{\sigma}}
\newcommand{\variance}{\mathbb{V}\mathrm{ar}}
\newcommand{\skewness}{\mathbb{S}\mathrm{kew}}
\newcommand{\exkurt}{\mathbb{E}\mathrm{x}\mathbb{K}\mathrm{urt}}

   \title{3D Full Spectrum Fitting: Algorithm Comparison}

   \author{
        Prashin Jethwa\inst{1},
        Simon Hubmer\inst{2},
        Ronny Ramlau\inst{2,3}
        \and
        Glenn Van de Ven\inst{1},
        }

   \institute{Institute for Astronomy (IfA), University of Vienna,
   T\"urkenschanzstrasse 17, A-1180 Vienna, Austria\\
   \email{prashin.jethwa@univie.ac.at}
   \and
   Johannes Kepler University Linz, Institute of Industrial Mathematics, Altenbergerstraße 69, 4040 Linz, Austria
   \and
   Johann Radon Institute Linz, Altenbergerstraße 69, 4040 Linz, Austria\\
    }

   \date{\today}

% \abstract{}{}{}{}{}
% 5 {} token are mandatory

  \abstract
   {
   Full spectrum fitting is the prevailing method for extracting stellar kinematic and population measurements from 1D galaxy spectra.
   3D methods refer to analysis of Integral Field Spectroscopy (IFS) data where spatial and spectral dimensions are modelled simultaneously.
   While several 3D methods exist for modelling gas structures probed by a single emission like, there has been less investigation into the more computationally demanding problem of 3D full spectrum fitting for stellar recoveries.
   }
   {
   This work introduces and compares two algorithms for 3D full spectrum fitting: the Projected Nesterov Kaczmarz Reconstruction method (PNKR) and a version of the Bayes-LOSVD software which has been modified to account for spatial correlations.
   We aim to understand strengths and weaknesses of both algorithms and assess the impact of 3D methods for stellar inferences.
   }
   {
   We apply both recovery algorithms to a mock IFS data with a ground truth model resembling a counter-rotating galaxy, over a signal-to-noise ratio (SNR) range from 20-200.
   We evaluate the quality of the recoveries compared to the known ground truth.
   }
   {
   Accounting for spatial correlations in Bayes-LOSVD significantly improved the accuracy and precision of kinematic recoveries.
   3D modelling with PNKR did not provide any significant improvement over 1D fits however, for SNR>40, PNKR did recover the most accurate kinematics overall.
   Additionally, by modelling the \emph{joint} distribution over kinematics and populations, PNKR could successfully infer trends between these quantities e.g. inferring local metallicity-velocity trends, albeit with a significant bias on the absolute metallicity.
   This successful demonstration of joint kinematic-population analyses includes cases where the counter-rotating components are not spectroscopically well-resolved.
   }
   {
   Having demonstrated advantages of (i) 3D modelling with Bayes-LOSVD, and (ii) joint kinematic-population analyses with PNKR, we conclude that both methodological advances will prove useful for detecting and characterising stellar structures from IFS data.
   }

   \maketitle

\section{Introduction}
\label{sec:intro}

The evolution of a galaxy is imprinted in its stellar structures.
By detecting and characterising these structures with increasing accuracy, we can make increasingly detailed inferences about a galaxy's history.
Integral Field Spectroscopy \citep[IFS,][]{Bacon95} allows us to map out stellar kinematics and populations in extragalactic systems.
The growing size and quality of IFS galaxy surveys \citep{califa,Bundy15,sami2,Sarzi18,GECKOS24} demand increasingly sophisticated methods to optimally measure these stellar properties.

Full-spectrum fitting \cite[e.g.][]{ppxf04,Tojeiro07,Koleva09,CidFernandes18} is the prevailing method to measure stellar kinematics and populations.
In this approach, the spectrum is modelled directly in pixel space using a wider wavelength range compared to earlier methods which focussed on a few equivalent widths or spectral indices \citep{Worthey94,Kauffmann03}.
Full-spectrum fitting typically combines single stellar population templates \citep[e.g.][]{BruzualCharlot03,Vazdekis10,Maraston20} with Line of Sight Velocity Distributions (LOSVDs) which are either described via Gauss Hermite expansions \citep{RixWhite92,SahaWilliams94} or non-parametrically \citep{RixWhite92,SahaWilliams94}.
These developments have cemented full spectrum fitting's position as a flexible and important tool for characterising stellar systems. However, one drawback to the method is that it treats the 3D datacube as a number of 1D spectra which are modelled independently.

3D methods refer to IFS analyses where spatial and spectral dimensions are modelled simultaneously.
This feature allows us to leverage spatial regularities during reconstructions to provide more accurate measurements.
3D modelling is an established technique for modelling gas components, with several tools available for modelling gas disks \citep{jozsa07,Bouche_15,barolo3d,Varidel19} alongside specialised tools for handling more complex structures such as bars \citep{davis13} and gravitationally lensed observations \citep{rizzo18}.
The relative dearth of 3D methods for stellar reconstructions may be explained by the fact that this is a much more computationally demanding problem since, while gas components are characterised by single emission lines, information on the stellar content is spread throughout the full spectrum.

The Projected Nesterov Kaczmarz Reconstruction method \citep[PNKR,][]{Hinterer23} is the first algorithm developed for 3D full spectrum fitting.
PNKR leverages the mathematical structure of the forward model to efficiently reconstruct stellar properties from IFS data.
Additionally, PNKR models \emph{joint} distribution over populations and kinematics, in contrast to most full spectrum fitting codes which assume that, at a given position in the galaxy, kinematics and populations are independent.
This means that PNKR has the freedom to infer local relations between populations and kinematics.
Previously, these types of measurements have been made via decomposition techniques for galaxies with spectroscopically-resolved substructure, i.e., galaxies with large scale stellar counter-rotation \citep{Coccato11,Johnston13,Coccato13,Pizzella18}.
In this work, we will test PNKR using mock data experiments, and in particular explore its ability to make joint population-kinematic inferences.

This work also introduces a second approach for 3D full spectrum fitting based on Bayes LOSVD \citep{bayeslosvd}, which is one of a number of recent tools \citep{Mehrgan23,GasymovKatkov24} for non-parametric LOSVD recoveries.
The use of non-parametric LOSVDs has enabled the detection of weak kinematic signatures \citep{Mehrgan23} and has been shown to affect dynamical inferences \citep{Reiter25}, however the extra flexibility of these models introduces degeneracy into the recoveries.
Typically this is handled using regularisation in velocity space, e.g., Figure 7 of \citet{bayeslosvd} illustrates the effect of velocity regularisation on LOSVDs recovered from the archetypal counter-rotating galaxy NGC4550 \citep{Rubin4550}.
The bottom two rows of that figure show how velocity regularisation tends to oversmooth LOSVD bimodality, which led the authors to recommend non-regularised fits over regularised ones.
In this work, we forego regularisation in velocity space and instead explore the effect of \emph{spatial} regularisation on LOSVD recoveries.

The paper is organised as follows.
Section~\ref{sec:methods} describes the PNKR and Bayes LOSVD algorithms and how they incorporate 3D modelling.
Section~\ref{sec:mock_data} describes the generation of mock data, and Section~\ref{sec:results} presents the results of reconstructions on mock data and compares the performance of PNKR and Bayes LOSVD.
We discuss the implications and limitations of these results in Section~\ref{sec:discuss} then summarise our main conclusions in \ref{sec:concs}.

\section{Methods}
\label{sec:methods}

This section introduces methods to reconstruct stellar population and kinematic distributions from IFS datacubes.
Concretely, we aim to recover the joint distribution $f(\xv,v,z,t)$ of stars over 2D position $\xv$, line-of-sight velocity $v$, stellar metallicity $z$ and age $t$.
The forward model introduced by \citet{Hinterer23} to relate $f$ to the 3D datacube signal $\bar{y}(\xv,\lambda)$, at position $\xv$ and wavelength $\lambda$, is given by:
\begin{equation}
    \bar{y}(\xv,\lambda) = \int \int \int \frac{1}{1+v/c} S\left(\frac{\lambda}{1+v/c},z,t\right) f(\xv,v,z,t) \;\mathrm{d}t \;\mathrm{d}z \;\mathrm{d}v,
    \label{eqn:forward model}
\end{equation}
where the kernels $S(\lambda;z,t)$ are single stellar population (SSP) templates.
Equation~\eqref{eqn:forward model} represents an idealised model assuming that effects unrelated to the stellar light (e.g. dust absorption and gas emission) can be otherwise modelled or corrected for.

We investigate two reconstruction algorithms, Bayes-LOSVD and PNKR.
While the former is limited to LOSVD recovery, PNKR models the full joint density $f(\xv,v,z,t)$ over position, velocity and populations.
Both algorithms have 1D versions, where spectra at different locations are fit independently (PNKR-1D \& BLOSVD-1D) as well as versions which fit multiple spectra simultaneously accounting for spatial correlations (PNKR-3D \& BLOSVD-2D).
The following sections introduce both algorithms, and demonstrate their 1D versions on a toy problem of recovering a Gaussian ground truth LOSVD.
For this toy problem, instead of SSP templates we use a single template spectrum consisting of a single absorption line, and mock data with signal to noise ratio (SNR) 50.
Applications to a more realistic scenario are given in Section~\ref{sec:results}.

\subsection{PNKR}
\label{ssec:pnkr}

The Projected Nesterov Kaczmarz Reconstruction method \citep[PNKR, ][henceforth \citetalias{Hinterer23}]{Hinterer23} was the first algorithm developed for 3D full spectrum fitting.
Given an observed datacube and set of SSP templates, PNKR inverts equation~\eqref{eqn:forward model} to return an estimate of the full joint density $f(\xv,v,z,t)$, which is described non-parametrically as a 5D array of weights.
PNKR is an \emph{iterative regularisation method}.
This means the density $f$ is iteratively updated, with later iterations adding increasingly detailed refinements.
By terminating the iterations at a suitable time, the solution should be regularised so as to avoid spurious details which are driven by over-fitting to noise.
A full mathematical description of the algorithm is given in \citetalias{Hinterer23}; the remainder of this section provides a qualitative overview.

\begin{figure*}
  \centering
  \resizebox{\hsize}{!}{\includegraphics{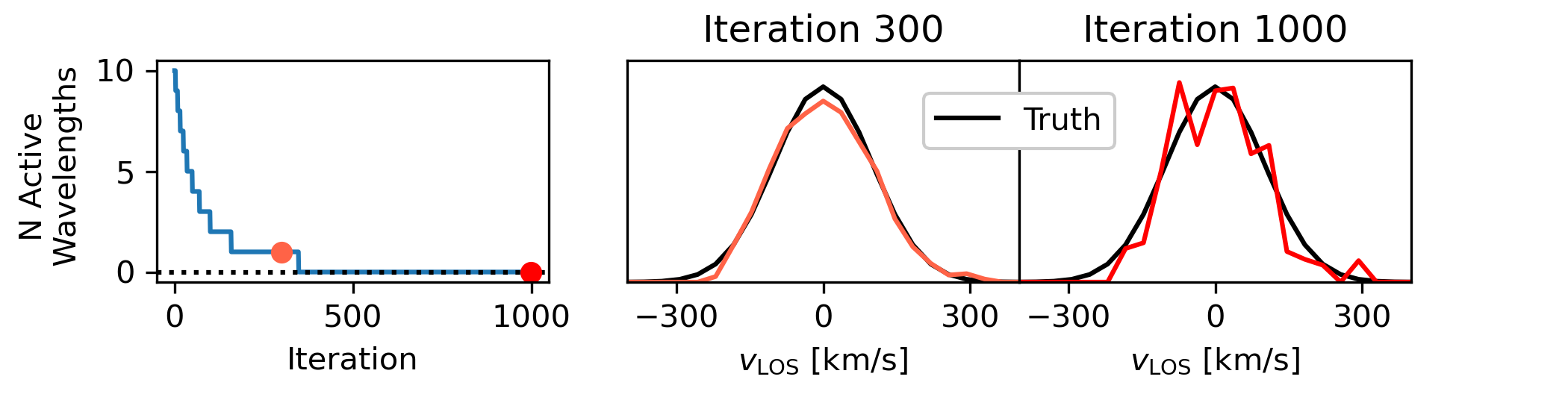}}
  \caption{
  Toy demonstration of LOSVD recovery with PNKR.
  The left panel shows the number of active wavelengths per iteration of the algorithm, with iterations 300 and 1000 highlighted.
  Corresponding LOSVD recoveries are shown in the right panels: at iteration 300, the recovery is smooth and matches the truth well, while extending to iteration 1000 introduces noise.
  }
  \label{fig:toy_pnkr}
\end{figure*}

PNKR's iteration scheme proceeds as follows.
The density is initialised as $f_0=0$, and all wavelengths $\lambda$ in the observed datacube are marked as \emph{active}.
In later iterations, PNKR cycles through all active wavelengths $\lambda_k$ and applies an update to the density $f_i$ which is a linear function of the datacube slice $y(\xv,\lambda_k$) and a pre-defined step-size.
If the reconstructed slice $y(\xv,\lambda_k$) fits the observed datacube within a provided SNR-dependant noise tolerance, the wavelength $\lambda_k$ is marked as \emph{inactive} then ignored in later iterations.
Iterations are continued until one of two stopping criteria is achieved: either the number of active wavelengths reaches zero or has plateaued within some tolerance.
Figure~\ref{fig:toy_pnkr} illustrates PNKR's iterative regularisation scheme on the toy problem.

The name Projected Nesterov Kaczmarz Reconstruction describes the following aspects of the algorithm.
After each iteration, the solution is projected to non-negative values to satisfy the physical constraint that $f\geq 0$.
The algorithm employs Nesterov acceleration \citep{Nesterov_1983} to speed up convergence.
Lastly, PNKR is an example of a Kaczmarz method \cite{Engl_Hanke_Neubauer_1996,Kaltenbacher_Neubauer_Scherzer_2008,Kindermann_Leitao_2014}, which are a class of methods for iteratively solving large linear problems with a specific system structure.
PNKR is a Kaczmarz method by virtue of the fact that we cycle through wavelengths at each iteration, which has the benefit that each update comprises a relatively cheap computation.

PNKR can be run in two modes: PNKR-1D, which is equivalent to independent spectral fits in each spaxel, and PNKR-3D, which accounts for spatial correlations during the reconstructions.
PNKR-1D follows the iterative scheme described above, which imposes no explicit coupling between adjacent spaxels in the recovery.
PNKR-3D incorporates spatial correlations in the theoretical framework of Adjoint Sobolev Embedding Operators \citep{Hubmer_Sherina_Ramlau_2023}.
For our problem, this means that we filter the density $f$ with a spatial smoothing kernel between iterations.
The smoothing kernel $K(\xv)$ is parametrised to allow us to explore different degrees of spatial smoothing.
It takes a shape parameter $\alpha$ and a dimensionless scale length $\beta$.
Roughly, these can be interpreted as follows: when $\alpha=1 (2)$ the kernel is Laplacian (Gaussian), and when $\beta=1$ the kernel scale is comparable to the full spatial domain.
In practice, the kernel is defined in Fourier space via
\begin{equation}
    \mathcal{F}_{\xv} \{K(\xv)\} (k) = \left(1 + \frac{\beta}{2} \left|\left|\frac{k}{k_\mathrm{max}}\right|\right|^2_2\right)^{-\alpha},
\end{equation}
where $k_\mathrm{max}=1/\Delta x$ and $\Delta x$ is the spaxel width.
The convolution of $f$ with $K$ is then numerically performed in Fourier space using Fast Fourier Transforms.

Lastly, we note that we have made one modification to the original \citetalias{Hinterer23} algorithm in order to account for a spatially varying SNR.
Specifically, we provide the algorithm with the true noise level used to generate our mock data - i.e. $\sigma(\xv,\lambda)$ from equation \eqref{eqn:noise_model} - which PNKR uses as a weighting factor when computing the overall residual between reconstructed and observed datacube slices.
Note that for real IFS data, standard data reduction pipelines \citep[e.g.][]{Weilbacher20} provide an estimate of this noise level.

\subsection{Bayes-LOSVD}
\label{ssec:meth_blosvd}

Bayes-LOSVD \citep[][henceforth \citetalias{bayeslosvd}]{bayeslosvd} is a framework for probabilistic recoveries of non-parametric LOSVDs from 1D spectral fits.
In this framework, two assumptions are imposed on equation \eqref{eqn:forward model} to reduce the problem size.
Firstly, in common with most full spectral fitting codes, \citetalias{bayeslosvd} assume that at a given position $\xv$, all stellar populations share the same kinematics.
This is equivalent to factorising the 5D density $f$ as
\begin{equation}
    f(\xv,v,z,t) = f_1(\xv) f_2(v|\xv) f_3(z,t|\xv),
\label{eqn:simplifying_assumption}
\end{equation}
which says that, locally (i.e. at a given $\xv$), the LOSVD ($f_2$) and stellar population distribution ($f_3$) are independent.
Secondly, \citetalias{bayeslosvd} replace SSP templates, of which there are typically several hundred, with a few templates distilled from the SSP library via Principal Component Analysis (PCA).
These two modifications reduce the size of the problem so that it is amenable to posterior sampling via Markov Chain Monte Carlo (MCMC) algorithms.
From a 1D spectral fit at position $\xv$, the main outputs of Bayes-LOSVD are posterior summaries (median and credible intervals) of the LOSVD $f_2(v|\xv)$, which is described non-parametrically as an array of weights in evenly-spaced velocity bins.

\begin{figure*}
  \centering
  \resizebox{\hsize}{!}{\includegraphics{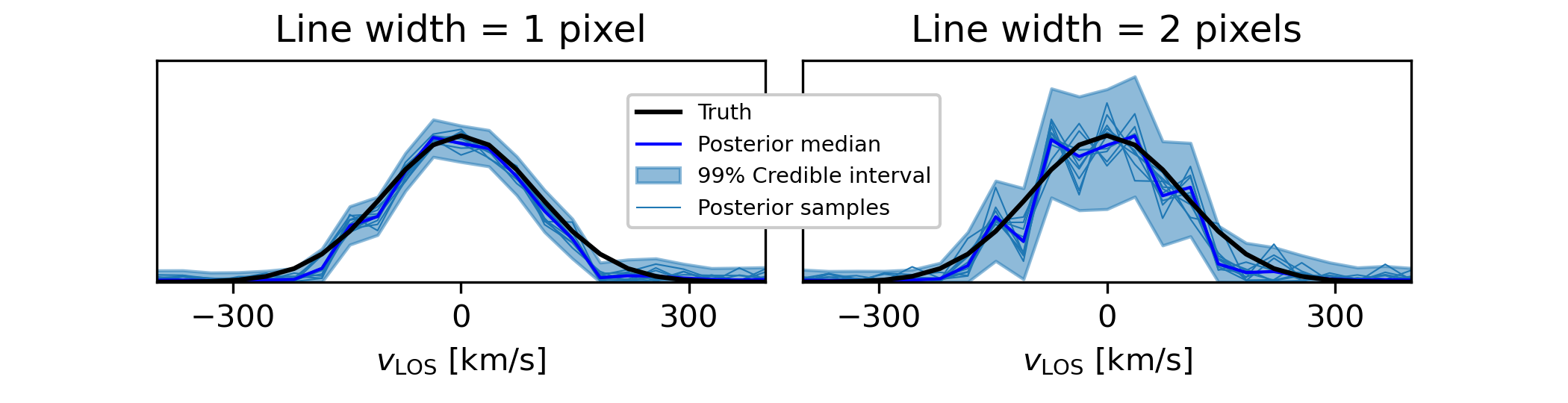}}
  \caption{
  Demonstration of LOSVD recovery with BLOSVD-1D.
  Each panel shows BLOSVD-1D recoveries via their posterior median (thick blue line), 99\% credible interval (shaded region) and 10 randomly selected posterior samples (thin blue lines).
  For the left panel, the template spectrum contains a single absorption line exactly one pixel wide; here, the recovery tightly encloses the truth.
  For the right panel, the absorption line is two pixels wide, which introduces a degeneracy between neighbouring velocity bins, evidenced by the \emph{zig-zagging} posterior samples.
  This degeneracy results in noisy median recoveries and inflated credible intervals.
  }
  \label{fig:toy_bayes}
\end{figure*}

Bayes-LOSVD provides a number of different options for regularisation of the LOSVD in velocity space.
The need for regularisation is illustrated in Figure~\ref{fig:toy_bayes}, which shows un-regularised LOSVD recoveries on the toy problem.
While the LOSVD is well recovered when the template spectrum has a one-pixel wide absorption line (left panel), when the line width is two pixels (right panel) noise is amplified and the credible intervals grow significantly.
At resolutions of interest for galactic astronomy, absorption lines in SSP templates are typically resolved into several pixels, hence un-regularised LOSVD recoveries are prone to exhibit the \emph{spike} artifacts seen in the right panel of Figure~\ref{fig:toy_bayes}.
Regularisation in velocity space offers one way to manage this, however it can oversmooth genuine kinematic substructures as discussed in Section \ref{sec:intro}.
We therefore explore spatial regularisation as an alternative to velocity regularisation.

We implement spatial regularisation in Bayes-LOSVD using a Conditional Autoregressive (CAR) prior \citep{car_besag74}.
CAR priors are used in Bayesian hierarchical models of spatial data \citep{Morris19}.
They encode the prior belief that adjacent sites in the model are normally distributed with respect to one another with some degree of correlation.
Mathematically, this is related to the spatial smoothing in PNKR for a Gaussian kernel.
For our application, we use a CAR prior to promote spatial smoothness of the LOSVD.
Specifically, if we denote the LOSVD value in velocity bin $i$ at spatial bin $j$ as $y_{ij} := f_2(v_i|x_j)$, and let $\left<y_i\right>_{\mathrm{Nbrs}(j)}$ be the mean LOSVD value in all spatial bins neighbouring $j$, then the CAR prior encodes that
\begin{equation}
    p(y_{ij} | \left<y_i\right>_{\mathrm{Nbrs}(j)},\rho_\mathrm{CAR},\sigma_\mathrm{CAR}) = \mathcal{N}\left(
        \rho_\mathrm{CAR}
        \left<y_i\right>_{\mathrm{Nbrs}(j)},
        \sigma_\mathrm{CAR}^2
    \right)
\label{eqn:car_prior}
\end{equation}
where the two hyper-parameters are a correlation $0<\rho_\mathrm{CAR}<1$ and a scale $\sigma_\mathrm{CAR}>0$.
The full CAR prior extends equation~\eqref{eqn:car_prior} to the joint distribution for all spatial bins using an efficient sparse matrix implementation to encode adjacency relations\footnote{See \url{https://mc-stan.org/learn-stan/case-studies/mbjoseph-CARStan.html} for details.}
We use CAR priors to spatially smooth the LOSVD in each velocity bin $i$, but use the same hyper-parameters $(\rho_\mathrm{CAR},\sigma_\mathrm{CAR})$ across all $i$ for simplicity.

To keep problem sizes manageable for these initial experiments, we do not use the CAR prior across the full cube, but instead restrict it's use to columns of spaxels.
We therefore denote this method BLOSVD-2D, rather than 3D.
Note that the choice of restricting to columns is for convenience, but is not a fundamental limitation to this method e.g. it would be straightforward to replace columns with clusters of adjacent spaxels.

We newly re-implement Bayes-LOSVD with modes for 1D fitting (BLOSVD-1D) and 2D fitting (BLOSVD-2D) using the probabilistic programming language \texttt{NumPyro} \citep{numpyro}.
This provides speed-ups of factor $\approx 10$ compared to the original \texttt{Stan} \citep{Stan} implementation used by \citetalias{bayeslosvd}.
In keeping with \citetalias{bayeslosvd}, for sampling we use the No-U-Turn Sampler \citep[NUTS,][]{NUTS} algorithm.
This is a gradient-based Markov-Chain Monte-Carlo (MCMC) algorithm which employs a warm-up stage for calibration followed by sampling stage during which posterior samples are efficiently drawn.
We check convergence via the standard NUTS diagnostics \citep{stanref}: the potential scale reduction factor $\hat{R}$, 
effective sample size $N_\mathrm{eff}$ and number of divergent MCMC transitions $N_\mathrm{div}$.
Unless explicitly mentioned, all Bayes-LOSVD results shown henceforth are converged subject to the criteria that all parameters satisfy $|\hat{R}-1|<0.03$, $N_\mathrm{eff}>50$ and $N_\mathrm{div}=0$.

\section{Mock data}
\label{sec:mock_data}

This section outlines the generation of mock datasets used to test our reconstruction algorithms.
In brief, we use simple analytic models to describe a ground truth model $f(\xv,v,z,t)$, evaluate the resulting datacube $y(\xv,\lambda)$ via equation~\eqref{eqn:forward model}, then add noise to achieve a desired SNR.
We do this all using the \texttt{popkinmocks} software \citep{Jethwa2023}
\footnote{{An example notebook is provided at \url{https://github.com/prashjet/popkinmocks/blob/main/notebooks/make_mock_data.ipynb} }}.

\begin{figure*}
  \centering
  \resizebox{\hsize}{!}{\includegraphics{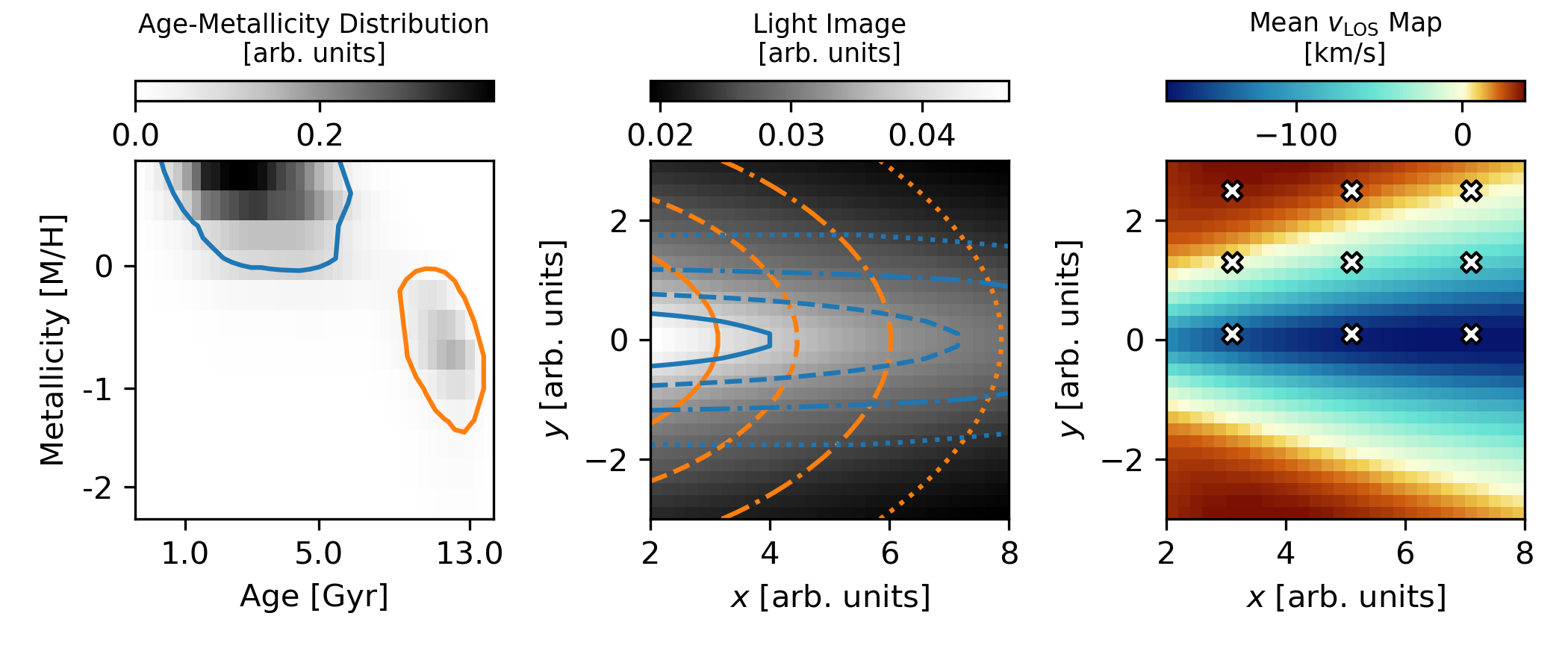}}
  \caption{
  Our ground truth galaxy model in age-metallicity space (left panel), as a light-weighted image (center) and mean velocity map (right).
  The model consist of two components.
  The dominant component (blue contours) is a young, metal-rich thin disk with negative LOS velocity.
  The counter-rotating weaker component (orange contours) is older, more metal-poor and more vertically extended.
  Successive contours in the central panel show changes in flux by 15\%.
  The grid of crosses in the right panel indicate spaxels used for illustration in Figures \ref{fig:pnkr_vs_blosvd}, \ref{fig:b3d}, and \ref{fig:metal_map}.
  }
  \label{fig:default_mock_data}
\end{figure*}

Figure~\ref{fig:default_mock_data} illustrates the default ground truth model used throughout this work.
The model consists of two counter-rotating disk components: a young, metal-rich thin disk (blue contours) and an older, metal-poor thick disk (orange contours).
The superposition of these two components gives rise to a mean velocity map (right panel) which exhibits an inversion, with negative velocities along the major axis swapping to positive velocities in the upper/lower-left corners.
The central/right panels of Figure~\ref{fig:default_mock_data} illustrate the field-of-view and spatial sampling used for our mock observations: we simulate one side of the galaxy, sampled with a $30\times30$ grid of spaxels.

Given this ground truth model $f(\xv,v,z,t)$, we next evaluate the noise free datacube $\bar{y}(\xv,\lambda)$ according to equation~\eqref{eqn:forward model}.
For this we adopt MILES SSPs \citep{Vazdekis_SanchezBlazquez_FalconBarroso_Cenarro_Beasley_Cardiel_Gorgas_Peletier_2010} with BaSTI isochrones for base alpha-content \citep{Pietrinferni04} and a Chabrier IMF \citep{Chabrier03}, using 12 metallicity bins spanning $-2.51\leq[M/H]\leq0.47$ and 38 age bins spanning 0.45-14.25 Gyr, and wavelength range 4800 - 5700 \r{A}.
For velocity sampling, we adopt a range of [-750, 750] km/s, fully covering the ground truth model, divided into 41 bins of width $\approx$ 37 km/s.
The \texttt{popkinmocks} software re-samples the SSPs in the equivalent $\log \lambda$, and evaluates equation~\eqref{eqn:forward model} using Fast Fourier Transforms to perform convolution of SSPs with LOSVDs.
This gives the noise-free datacube $\bar{y}(\xv,\lambda)$.

To arrive at the observed datacube $y(\xv,\lambda)$, we add noise
\begin{equation}
    y(\xv,\lambda) = \bar{y}(\xv,\lambda) + \epsilon(\xv,\lambda)
\end{equation}
which we sample noise from a normal distribution with mean 0 and variance $\sigma^2(\xv,\lambda)$ i.e.
\begin{equation}
    \epsilon(\xv,\lambda) \sim \mathcal{N}\left(0, \sigma^2(\xv,\lambda) \right)
    \label{eqn:noise_model}
\end{equation}
We take $\sigma^2(\xv,\lambda)$ proportional to $\bar{y}(\xv,\lambda)$ to mimic shot noise and pick a constant of proportionality in order to achieve a desired SNR in the brightest spaxel.
Throughout this work, we investigate five SNR values logarithmically spaced between SNR=20 and 200.
These SNR values refer to the the brightest spaxel; the noise model described above gives rise to spatial variations in SNR of around 30\%.
We do not apply any line-spread function beyond that inherent to the SSP templates nor apply any point-spread function or other sources of noise.

\section{Results}
\label{sec:results}

This section presents reconstructions using the mock data sets described in Section~\ref{sec:mock_data}.
Sections \ref{ssec:1d_results} and \ref{ssec:23d_results} focus purely on LOSVD reconstructions, first comparing the 1D versions of PNKR and BLOSVD (Section~\ref{ssec:1d_results}), then their 2D/3D equivalents (Section~\ref{ssec:23d_results}).
In Section~\ref{ssec:popkin_pnkr} we explore PNKR's ability to recover the \emph{joint} distribution over stellar populations and kinematics.
Additional details on implementation and fit convergence are given in Appendix~\ref{apdx:implementation} and spectral fits are shown in Appendix~\ref{apdx:qof}.

Before presenting results, we quantify the size of the reconstruction problems we are describing.
We aim to reconstruct an unknown density $f(\xv,v,z,t)$ with sampling in spatial, velocity and metallicity dimensions as described in Section~\ref{sec:mock_data}.
In order to keep run-times for PNKR within a reasonable range ($\lesssim$ 48 hours), for reconstructions we chose to thin the age sampling of the SSP grid by a factor 3, reducing the number of ages from 38 to 13.
Nevertheless, in total, we sample $f(\xv,v,z,t)$ with $30 \cdot 30 \cdot 41 \cdot 12 \cdot 13 \approx 6\times10^6$ unknown entries.
The simplifying assumptions of Bayes-LOSVD reduce this number significantly however we stress that PNKR simultaneously solves for $6\times10^6$ unknowns, constituting a significantly high dimensional problem.

\subsection{LOSVD recovery from 1D fits}
\label{ssec:1d_results}

% PNKR-1D settings described in Section~\ref{ssec:pnkr}.
% Particular choices we used: stepsize 100, tau 1.3 i.e. if a wavelength is fit to 30\% accuracy it is deactivated.
% 1D fits, all fits reach 0 active equations before the algorithm terminates.
% Number of iterations required to reach 0 active equations decreases with SNR: 560 (SNR 200) to 74 (SNR 20).
% Time to finish: 15 hours - 1 hour.

% BLOSVD-1D settings: number of PCA templates (15).
% HMC settings: 1000 warmup steps, 2000 sampling steps.
% Convergence checks: all rhat < 1.1, all Neff > 100.
% Fits terminate successfully for all SNR. 
% Time to finish 900 spaxels: 30 minutes (high SNR) - (3 minutes) low SNR.
% Why much faster than PNKR?
% Straightforward to parallelise: different spectra, different process.

\begin{figure*}
  \centering
  \includegraphics[width=\textwidth]{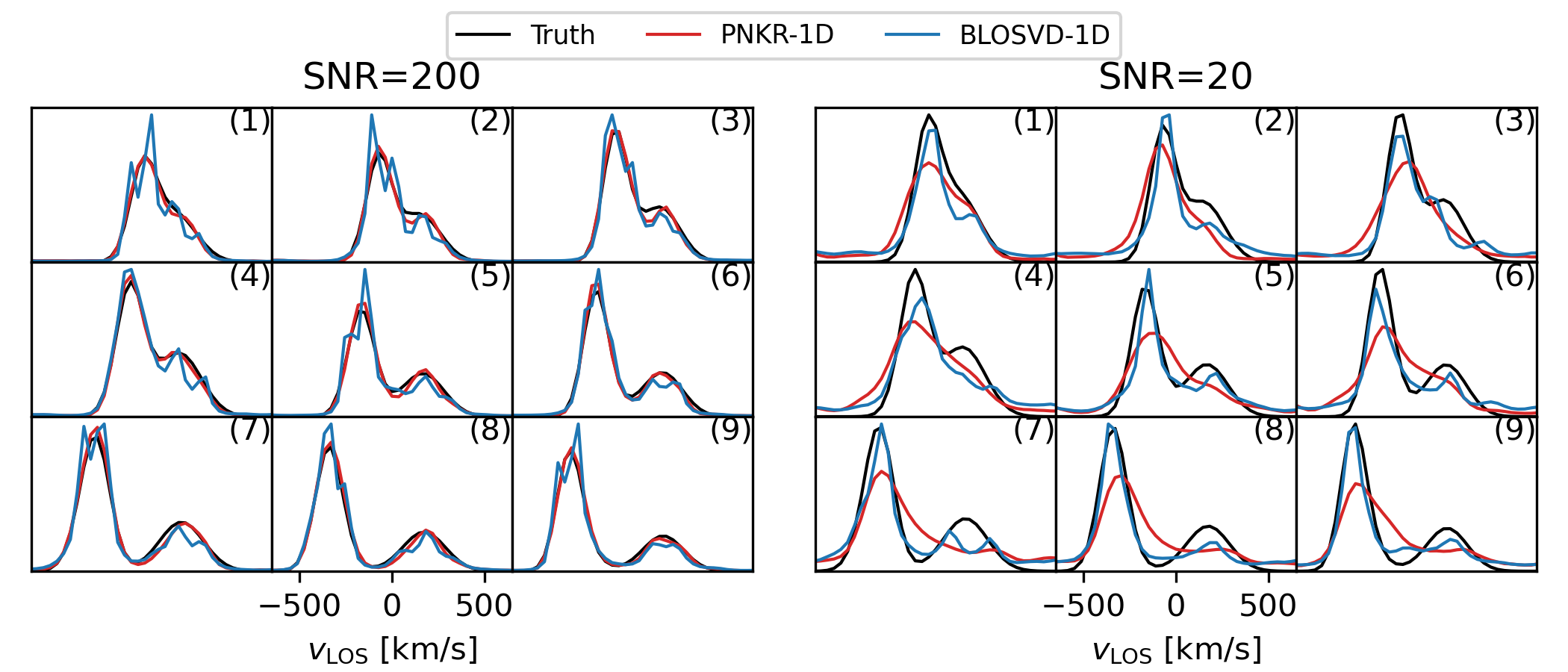}
  \caption{
  Comparison of LOSVDs recovered from 1D spectral fits.
  True LOSVDs (black) are shown alongside recoveries from PNKR-1D (red) and BLOSVD-1D (blue).
  At SNR 200 (left grid) PNKR recovers the LOSVD well everywhere, while BLOSVD recoveries show \emph{spike} artifacts in some cases e.g. panels 1, 2 and 9.
  At SNR 20 (right grid) the recoveries are worse.
  Both algorithms produce LOSVDs with extended, flat wings which reach the edge of the velocity range.
  For spaxels with bimodal LOSVDs (panels 5-9) we see that BLOSVD-1D captures the dip between the two components whereas PNKR-1D oversmooths this feature.
  }
  \label{fig:pnkr_vs_blosvd}
\end{figure*}

Figure~\ref{fig:pnkr_vs_blosvd} shows the recovery of light-weighted LOSVDs (defined in Appendix~\ref{apdx:lw_losvds}) at 9 spaxels whose locations are shown by the crosses in Figure~\ref{fig:default_mock_data}.
In spaxels 3-9, the two-component galaxy model gives rise to bimodal LOSVDs while in spaxels 1 and 2 the components overlap in $v_\mathrm{LOS}$ leading to unimodal LOSVDs.
At SNR 200 (left grid), PNKR-1D recoveries (red lines) are good everywhere, showing only minor deviations from the truth e.g. at $v_\mathrm{LOS}=0$ in spaxel 3.
The PNKR-1D recoveries are also smooth, suggesting that the algorithm's iterative regularisation scheme has worked well.
In contrast, the BLOSVD-1D recoveries (blue lines, showing posterior medians) at SNR=200 show large spike artifacts similar to those seen in Figure~\ref{fig:toy_bayes}.
At SNR 20 (right grid), both sets of recoveries are worse.
Both algorithms produce erroneous flat wings extending to the edge of the velocity range.
Additionally, PNKR-1D recoveries are over-smoothed in many spaxels (e.g. 7-9) where BLOSVD-1D does manage to recover the strong dip between the two components.
At this low SNR, BLOSVD-1D no longer produces spike artifacts, nevertheless the recovered LOSVD shapes are visibly worse than high SNR.

\begin{figure}
  \centering
  \includegraphics{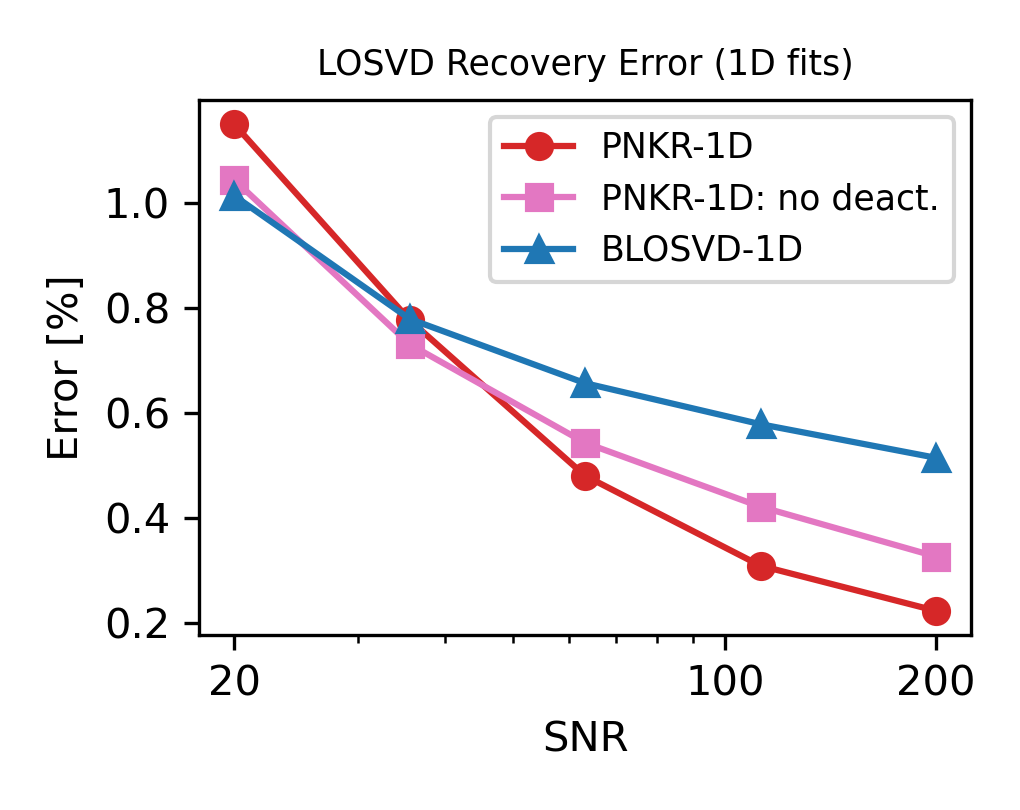}
  \caption{
  Error in recovered LOSVDs from 1D spectral fits as a function of SNR.
  At high SNR, PNKR (red) achieves the smallest error, while BLOSVD (blue) is best at low SNR.
  By modifying PNKR to retain all wavelengths throughout the recovery (pink line), we see an improvement at low SNR, with a recovery error similar to BLOSVD, at the expense of larger errors at high SNR.
  }
  \label{fig:1d_summary}
\end{figure}

Figure~\ref{fig:1d_summary} summarises the LOSVD recovery capabilities of the 1D algorithms.
It shows the the LOSVD recovery error averaged over all spaxels and velocity bins (defined in Appendix~\ref{apdx:losvd_rec_err}) as a function of SNR.
PNKR-1D (red line) achieves the lowest error for SNR>40 while BLOSVD-1D (blue line) performs best at SNR=20.
These results show a smooth transition between the two SNR regimes depicted in Figure~\ref{fig:pnkr_vs_blosvd}, where oversmoothing affects the PNKR-1D recoveries at low SNR and spike artifacts affect BLOSVD-1D at high SNR.

We next investigated why PNKR-1D tends to oversmooth at low SNR.
We varied the algorithm's step-size and tolerance settings (see Appendix~\ref{apdx:imp_bl1d} for details) but found this had no effect.
We then modified the algorithm itself. 
In the standard algorithm as described in Section~\ref{ssec:pnkr}, once a wavelength is well-fit it is \emph{deactivated} and not used to update the density $f$ in later iterations.
The pink line in Figure~\ref{fig:1d_summary} shows LOSVD recovery errors for a modified algorithm with \emph{no deactivation}.
This modification improves PNKR-1D reconstructions at SNR=20, bringing errors in line with BLOSVD-1D, however for SNR>60 we see that reconstructions become worse.
This test suggests that while wavelength deactivation benefits PNKR's iterative regularisation at high SNR it is likely connected to oversmoothing at low SNR.

\subsection{LOSVD recovery from 2D/3D fits}
\label{ssec:23d_results}

We next run reconstructions accounting for spatial correlations between spaxels.
We run 16 PNKR-3D recoveries over a ($5\times5$) grid of smoothing parameters $(\alpha,\beta)$ detailed in Appendix~\ref{apdx:imp_pnkr3d}.
For BLOSVD-2D, in principle it is possible to infer the values of the smoothing parameters $(\rho_\mathrm{CAR},\sigma_\mathrm{CAR})$ by assigning them suitable hyper-priors then sampling their posteriors alongside the LOSVD.
In practice, we find that this approach works well for the correlation parameter $\rho_\mathrm{CAR}$ but leads to problems when applied to the scale parameter $\sigma_\mathrm{CAR}$.
Specifically, MCMC chains exhibit many divergent transitions when they explore small values of $\sigma_\mathrm{CAR}$.
For BLOSVD-2D, we therefore run two reconstructions using two fixed values of $\sigma_\mathrm{CAR}=0.001, 0.03$, but retain the hyper-prior over $\rho_\mathrm{CAR}$.
Further details are given in Appendix~\ref{apdx:imp_bl3d}.
In this section, we show results for the best-performing set of smoothing parameters (as measured by LOSVD recover error), which are chosen independently at each SNR level.

\begin{figure}
  \centering
  \includegraphics{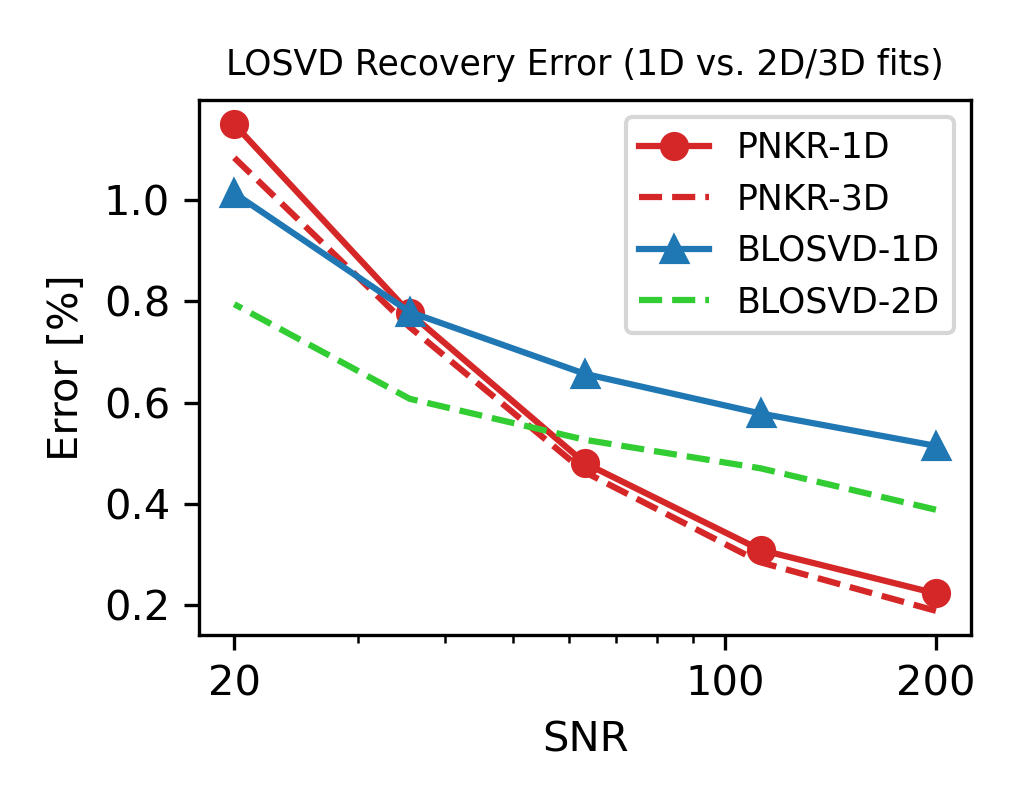}
  \caption{
  Error in recovered LOSVDs from 2D/3D fits as a function of SNR.
  The solid lines are repeated from Figure~\ref{fig:1d_summary}, showing the results of 1D spectral fits with PNKR (red) and BLOSVD (blue).
  The dashed lines show results when fitting multiple spaxels simultaneously and promoting spatial smoothness in the recoveries.
  PNKR-3D (red-dashed), which fits the full 3D datacube, produces results which are almost indistinguishable from the 1D PNKR fits.
  In contrast, BLOSVD-2D (green-dashed), which fits columns of spaxels simultaneously, achieves a significant improvement over the 1D case for all SNR values.
  }
  \label{fig:3d_summary}
\end{figure}

Figure~\ref{fig:3d_summary} shows LOSVD recovery errors for 2D/3D fits (dashed lines) compared to their 1D equivalents (solid lines, repeated from Figure~\ref{fig:1d_summary}).
We see that PNKR-3D produces very little improvement ($\approx 0.01\%$) compared to PNKR-1D at all SNR.
This result is not limited by the range of smoothing parameters we have explored (see Appendix~\ref{apdx:imp_pnkr3d}).
Furthermore, we have checked that this result does not change when (i) varying algorithm step-size and tolerance settings, (ii) initialising PNKR-3D fits not from zero, but instead from the outputs of the PNKR-1D fits, and (iii) using a smoothing filter based on wavelets, rather than Fourier convolution.
We conclude that PNKR-3D leads to very little improvement over PNKR-1D, however we also note that both versions of PNKR achieve lower reconstructions errors than BLOSVD for SNR>40. We will discuss these results further in Section~\ref{ssec:disc_PNKR}.

For BLOSVD, the impact of accounting for spatial correlations in LOSVD reconstructions looks very different.
In Figure~\ref{fig:3d_summary}, comparing 1D results (blue) to the 2D results (green), we see that BLOSVD-2D significantly outperforms BLOSVD-1D at all SNR, with an improvement of 0.3\% at SNR=20 and 0.1\% at SNR=200.
We devote the remainder of this section to a more detailed comparison of BLOSVD-1D vs BLOSVD-2D.

\begin{figure*}
  \centering
  \includegraphics[width=\textwidth]{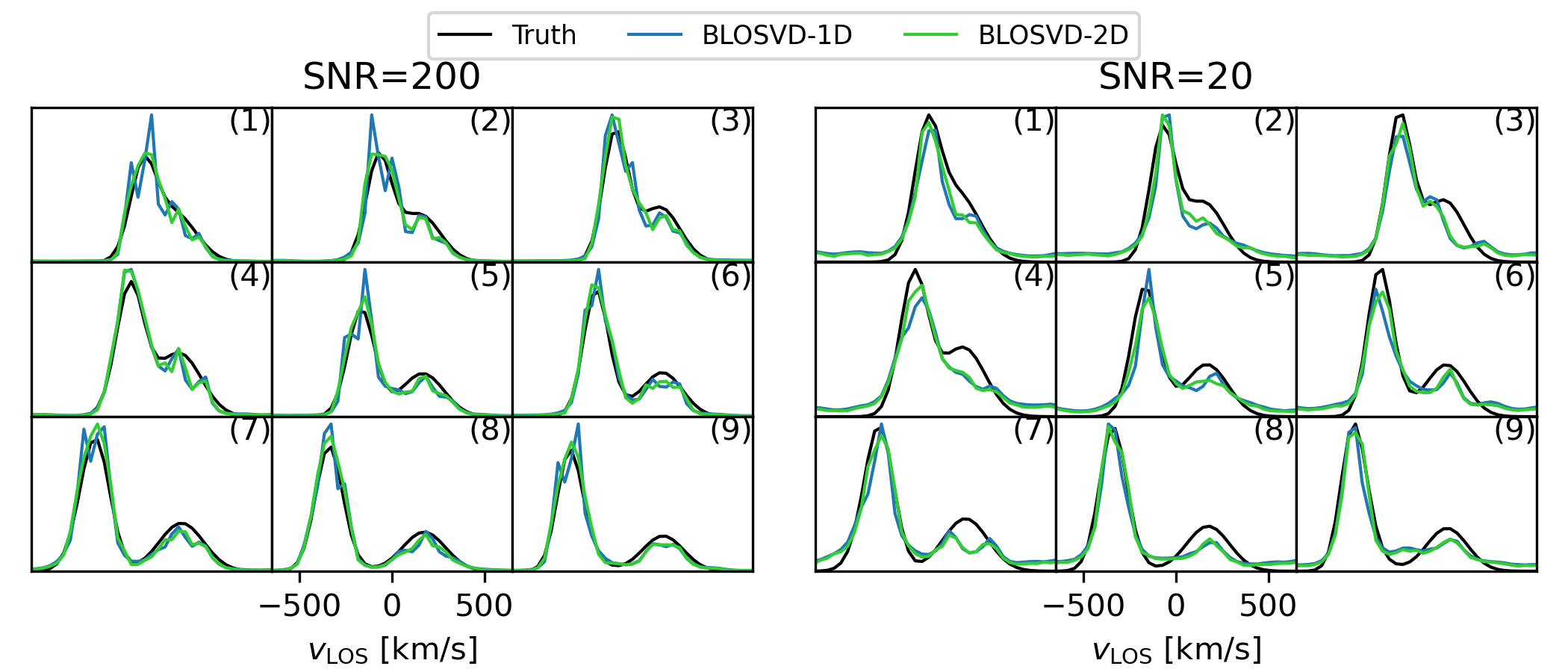}
  \caption{
  Comparison of LOSVDs recovered from with BLOSVD-1D (blue lines) against BLOSVD-2D (green lines).
  At SNR 200 (left grid), the spike artifacts seen in the 1D fits are alleviated in the 2D fits (e.g. panels 1, 2, 5, 9).
  At SNR 20 (right grid), in many cases 2D fits improve the recovered LOSVD shapes, particularly for the dominant component with $v_\mathrm{LOS}<0$ (e.g. panels 4, 6, 9).
  For both high and low SNR, the influence of 2D fitting is less strong for the weaker component with $v_\mathrm{LOS}>0$.
  }
  \label{fig:b3d}
\end{figure*}

Figure~\ref{fig:b3d} compares median LOSVD reconstructions for BLOSVD-1D (blue) and 2D (green) for 9 example spaxels.
At SNR=200 (left grid), we see that most severe spike artifacts are effectively regularised away when using BLOSVD-2D.
This is most evident for spikes occurring at the LOSVD maximum (e.g. spaxels 1, 2, 5), whereas spikes in the LOSVD of the subdominant component with $v_\mathrm{LOS}>0$ show little improvement (e.g. spaxels 2, 3, 4).
At SNR=20 (right grid), BLOSVD-2D recoveries show slight improvements in a few spaxels (e.g 4, 6, 9), again more so for the $v_\mathrm{LOS}<0$ component than the $v_\mathrm{LOS}>0$ one.
The flat wing artifacts in the SNR=20 recoveries are unchanged between 1D and 2D modelling.

\begin{figure*}
  \centering
  \includegraphics[width=0.95\textwidth]{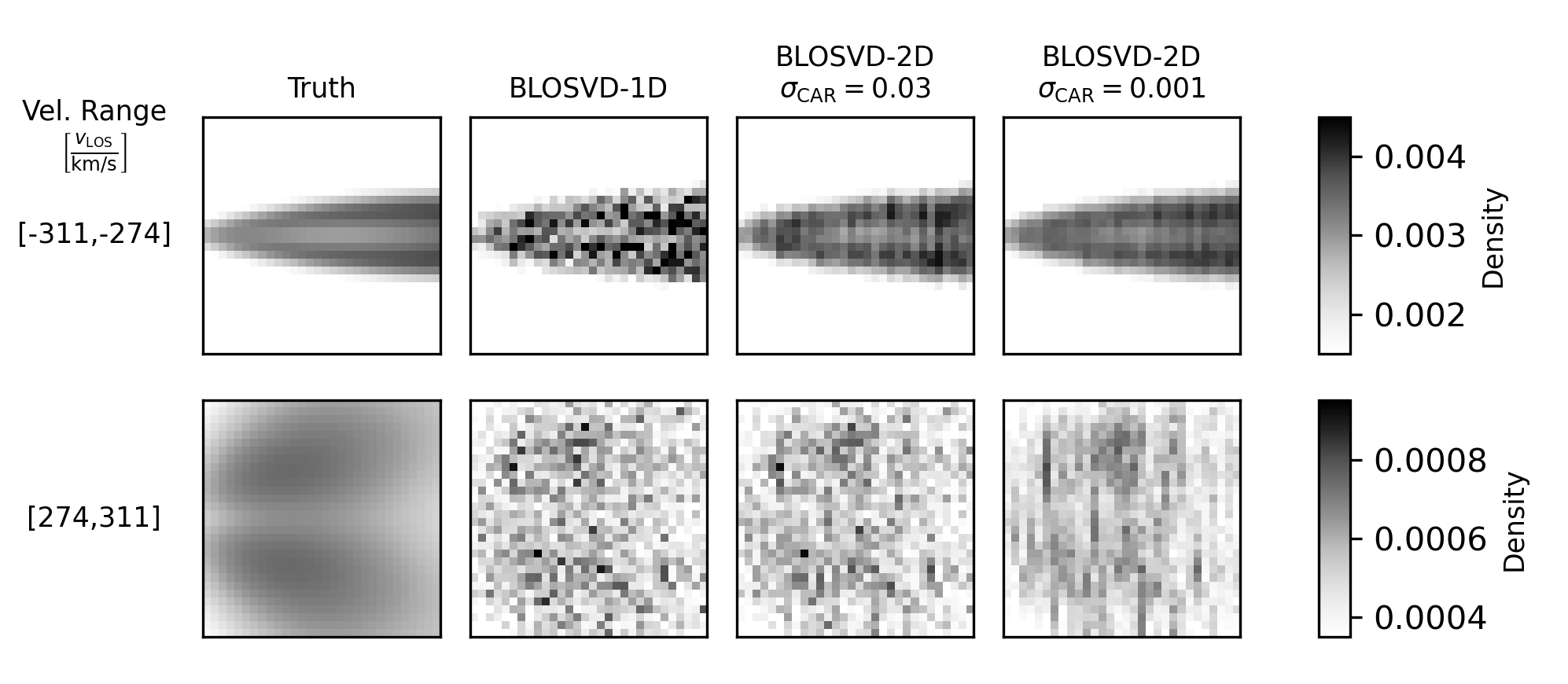}
  \caption{
  Channel maps showing the spatial distribution of stars in velocity bins labelled on the left.
  The first column shows true distributions which are smooth by construction.
  The second column shows BLOSVD-1D reconstructions, which show large spaxel-to-spaxel variations.
  The final two columns show BLOSVD-2D reconstructions for two choices of $\sigma_\mathrm{CAR}$.
  These display vertical stripes arising from spatially regularising over columns of spaxels.
  The degree of smoothing is greater for the dominant, negative velocity component (top row) compared to the positive velocity component (bottom), and also for $\sigma_\mathrm{CAR}=0.001$ (fourth column) versus $\sigma_\mathrm{CAR}=0.03$ (third).
  Spatial regularisation has proved most effective in the top-right panel, where $\sigma_\mathrm{CAR}=0.001$ is less than the density present in the channel map i.e. 0.002-0.004, as seen in the colour-bar.
  }
  \label{fig:channel_maps}
\end{figure*}

Figure~\ref{fig:channel_maps} compares BLOSVD-1D and 2D reconstructions at SNR=200 via channel maps.
These show the spatial distribution of stars in a given velocity bin, providing an intuitive way to visualise the effect of spatial regularisation on LOSVD reconstructions.
In general, we see that channel maps from 1D reconstructions (second column) are nosier than their 2D equivalents (third/fourth columns for $\sigma_\mathrm{CAR}=0.03/0.001$).
For the negative velocity channel (top row), corresponding to the dominant galactic component, BLOSVD-2D provides some smoothing for either choice of $\sigma_\mathrm{CAR}$, with $\sigma_\mathrm{CAR}=0.001$ (top-right panel) being the most accurate.
For the positive velocity channel (bottom row) the BLOSVD-2D reconstruction with $\sigma_\mathrm{CAR}=0.03$ provides little change with respect to the 1D case, while for $\sigma_\mathrm{CAR}=0.001$ the reconstructions are smoother but remain clumpier than the truth.
These results suggest that the effectiveness of spatial regularisation for a given velocity channel depends both on $\sigma_\mathrm{CAR}$ and the LOSVD density values in that channel.

\begin{figure*}
  \centering
  \includegraphics[width=\textwidth]{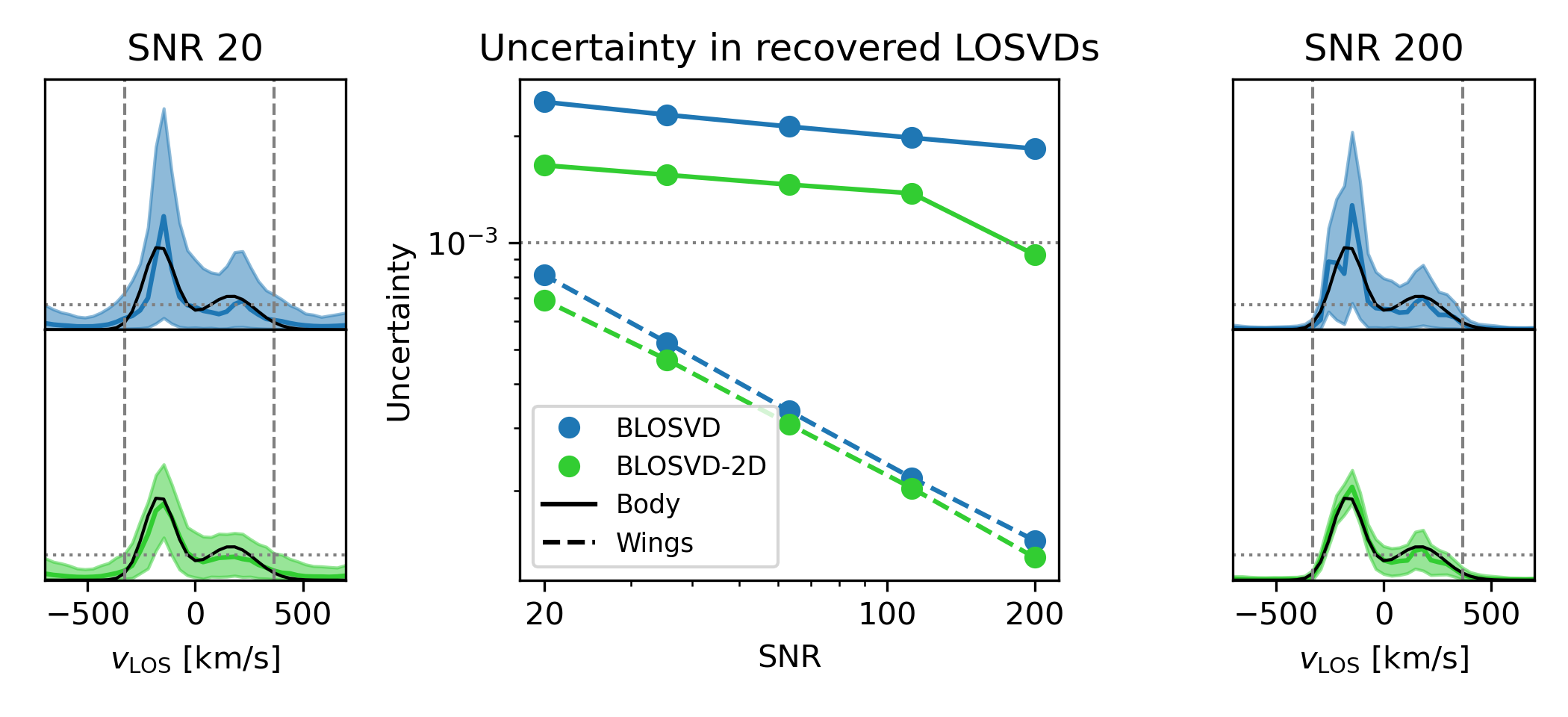}
  \caption{
  The uncertainty of LOSVDs recovered with BLOSVD (blue) and BLOSVD-2D (green).
  \textbf{Side panels} show an example from the central spaxel in Figure~\ref{fig:b3d}: the true LOSVD (black line) is compared against the median recovery (solid line) and 99\% credible intervals (shaded regions), for SNR 20 (left panels) and 200 (right).
  All four panels share the same $y$-axis range (a horizontal dotted line at $y=10^{-3}$ has been added to aid comparison).
  For 1D fits with BLOSVD (top left/right), the uncertainties are so large that almost all of the LOSVD body (i.e. within the vertical dashed lines) is consistent with zero; increasing from SNR 20 to 200 reduces uncertainty in the wings but much less so in the body.
  BLOSVD-2D, in contrast, gives credible intervals which enclose the truth more tightly: for SNR 20 (bottom left), the dominant negative-velocity component is recovered with high statistical significance, while for SNR 200 (bottom right) this is also true for the weaker component.
  \textbf{The central panel} shows the uncertainty (measured by the width of 99\% credible intervals) averaged over all spaxels and velocity bins, for a range of SNR.
  BLOSVD-2D shrinks uncertainties in the LOSVD body at all SNR (solid lines).
  In the wings (dashed lines) the difference is less stark, but varying SNR influences the uncertainties much more strongly.
  }
  \label{fig:b3d_uncertainties}
\end{figure*}

Figure~\ref{fig:b3d_uncertainties} compares uncertainties derived from BLOSVD-1D (blue) vs. BLOSVD-2D (green).
The example reconstructions at SNR=20 (left column) and SNR=200 (right) illustrate that BLOSVD-2D produces significantly smaller uncertainties which enclose the truth more tightly.
For SNR=20, we also see that the reduction in uncertainty is more prominent within the distribution body than the wings.
The central panel shows uncertainties averaged over all spaxels for a range of SNR.
In general, we see that BLOSVD-2D significantly decreases uncertainty in the LOSVD body at all SNR (solid lines).
The extra dip in the BLOSVD-2D uncertainty for SNR=200 arises from the choice of $\sigma_\mathrm{CAR}$. 
For SNR=200 we show results for $\sigma_\mathrm{CAR}=0.001$, however for SNR<200 this choice led to poor sampling diagnostics, so we instead show results for $\sigma_\mathrm{CAR}=0.03$.
For the LOSVD wings (dashed lines), accounting for spatial correlations decreases uncertainty much less than in the body, however increasing SNR has a much stronger effect.

\subsection{Recovery of Metallicity-Velocity Relations with PNKR}
\label{ssec:popkin_pnkr}

We now examine PNKR's ability to recover dependencies between stellar populations and kinematics encoded in the 5D joint distribution $f(\xv,v,z,t)$.
All results here use PNKR-1D; as with the LOSVD recoveries in Section~\ref{ssec:23d_results}, we found no improvement when using PNKR-3D.
This section focusses on reconstructions of metallicity-velocity relations; equivalent results on age-velocity relations are shown in Appendix~\ref{apdx:age_vel}.

\begin{figure*}
  \centering
  \includegraphics[width=\textwidth]{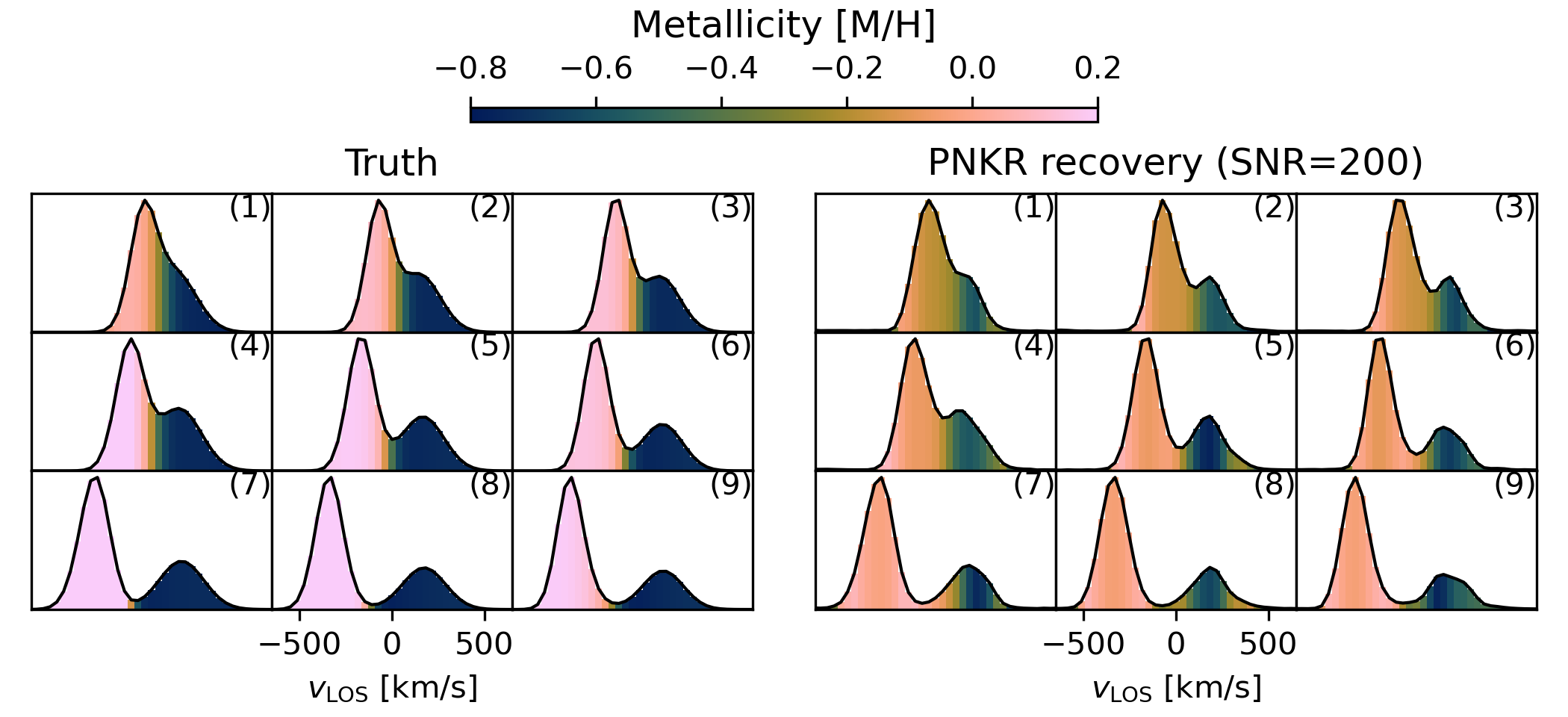}
  \caption{
  Recovery of metallicity-velocity relations with PNKR.
  LOSVDs are coloured by the mean metallicity of stars with a given $v_\mathrm{LOS}$, for the truth (left) and PNKR recovery at SNR 200 (right).
  Overall, PNKR successfully recovers that the negative-velocity component is more metal rich than the positive.
  This occurs where the two components are well separated in $v_\mathrm{LOS}$ (e.g. spaxels 7-9) but also where they overlap (spaxel 1).
  Though the metallicity ordering of the two components is successfully inferred, the absolute values of the recovered metallicities are less extreme than the true values, and some spaxels (e.g. 5 and 8) show an erroneous upturn in recovered metallicity at large positive $v_\mathrm{LOS}$.
  }
  \label{fig:metal_map}
\end{figure*}

Figure~\ref{fig:metal_map} shows PNKR recoveries of metallicity-velocity relations.
It shows LOSVDs from 9 spaxels coloured by the mean metallicity of stars at the given position and velocity (defined in Section~\ref{apdx:local_vel_met}), both for the ground-truth model (left grid) and the PNKR recovery at SNR=200 (right grid).
Overall, PNKR successfully recovers that the negative velocity component is more metal rich than the positive velocity component.
Even for spaxels where the two components overlap (e.g panel 1), and are therefore spectroscopically unresolved, we are able to recover a metallicity difference between the two components.
While we correctly infer the metallicity ordering of the two components, we see that the absolute values of the recovered metallicity are less extreme than the true values.
Additionally, some spaxels (e.g. 5, 8) show an erroneous upturn in metallicity at large positive $v_\mathrm{LOS}$.

\begin{figure}
  \centering
  \includegraphics{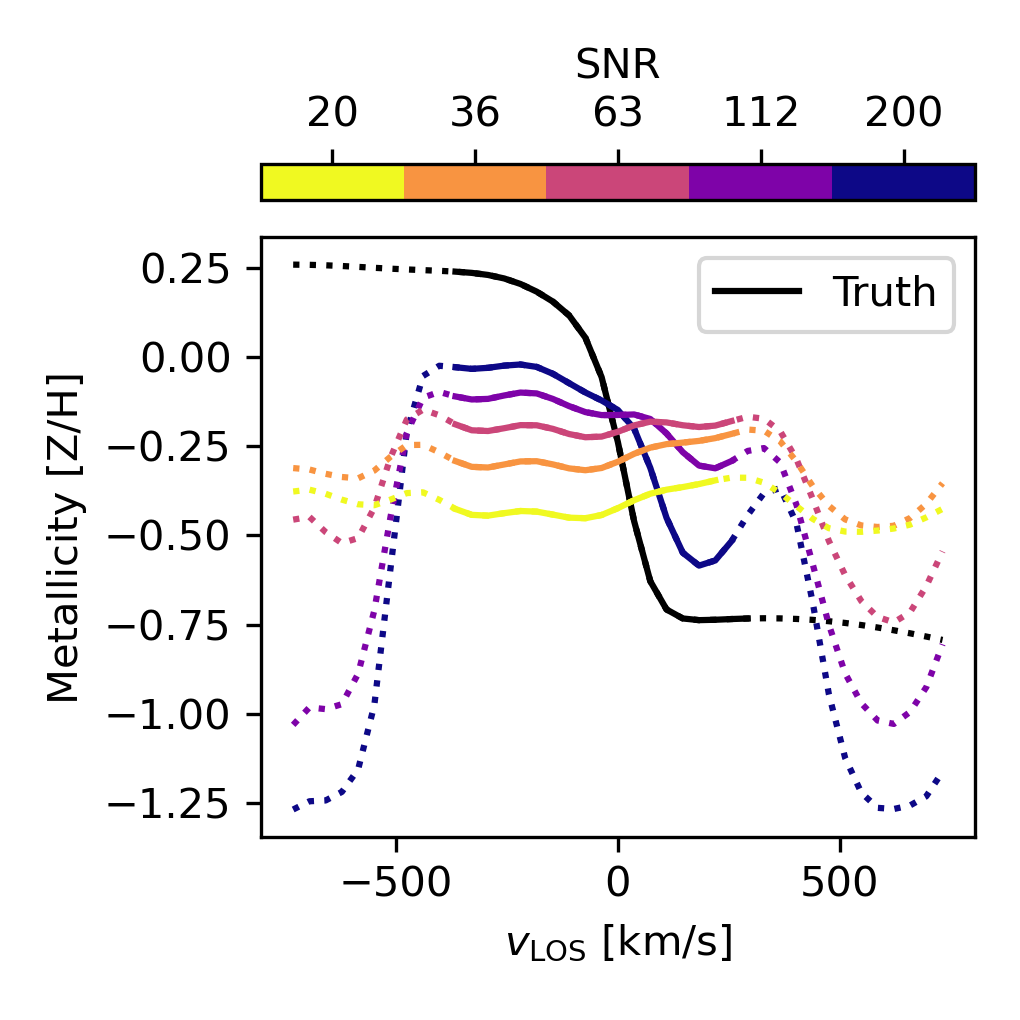}
  \caption{
  Spatially-integrated metallicity-velocity relations.
  The true relation (black line) is compared against recoveries at varying SNR (coloured lines).
  Dotted lines show velocities with a negligible contribution to the galaxy, where the metallicity is unconstrained.
  A sharp drop in metallicity is recovered for SNR=200, marginally so at SNR=112, but is washed out for lower SNR.
  }
  \label{fig:metal_snr}
\end{figure}

Figure~\ref{fig:metal_snr} shows spatially integrated metallicity-velocity relations (defined in Appendix~\ref{apdx:global_vel_met}) at different SNR levels.
Dotted lines indicate velocities which contribute negligibly to the density (selected via a LOSVD threshold of 0.01), where metallicities are unconstrained and we see recoveries diverge significantly from the truth.
Focussing instead on the solid lines, Figure~\ref{fig:metal_snr} shows that at SNR=200 (blue line) PNKR successfully recovers a sharp decline in metallicity but this becomes washed-out for lower SNR.
Though the qualitative behaviour of the metallicity-velocity relation is well-recovered at SNR=200, absolute metallicities are offset from the truths by 0.25/0.2 dex at negative/positive $v_\mathrm{LOS}$.
We defer discussion of these absolute offsets to Section~\ref{ssec:disc_PNKR}, and next explore whether PNKR can infer \emph{qualitative} behaviour of metallicity-velocity relations in different scenarios.

\begin{figure*}
  \centering
  \includegraphics[width=\textwidth]{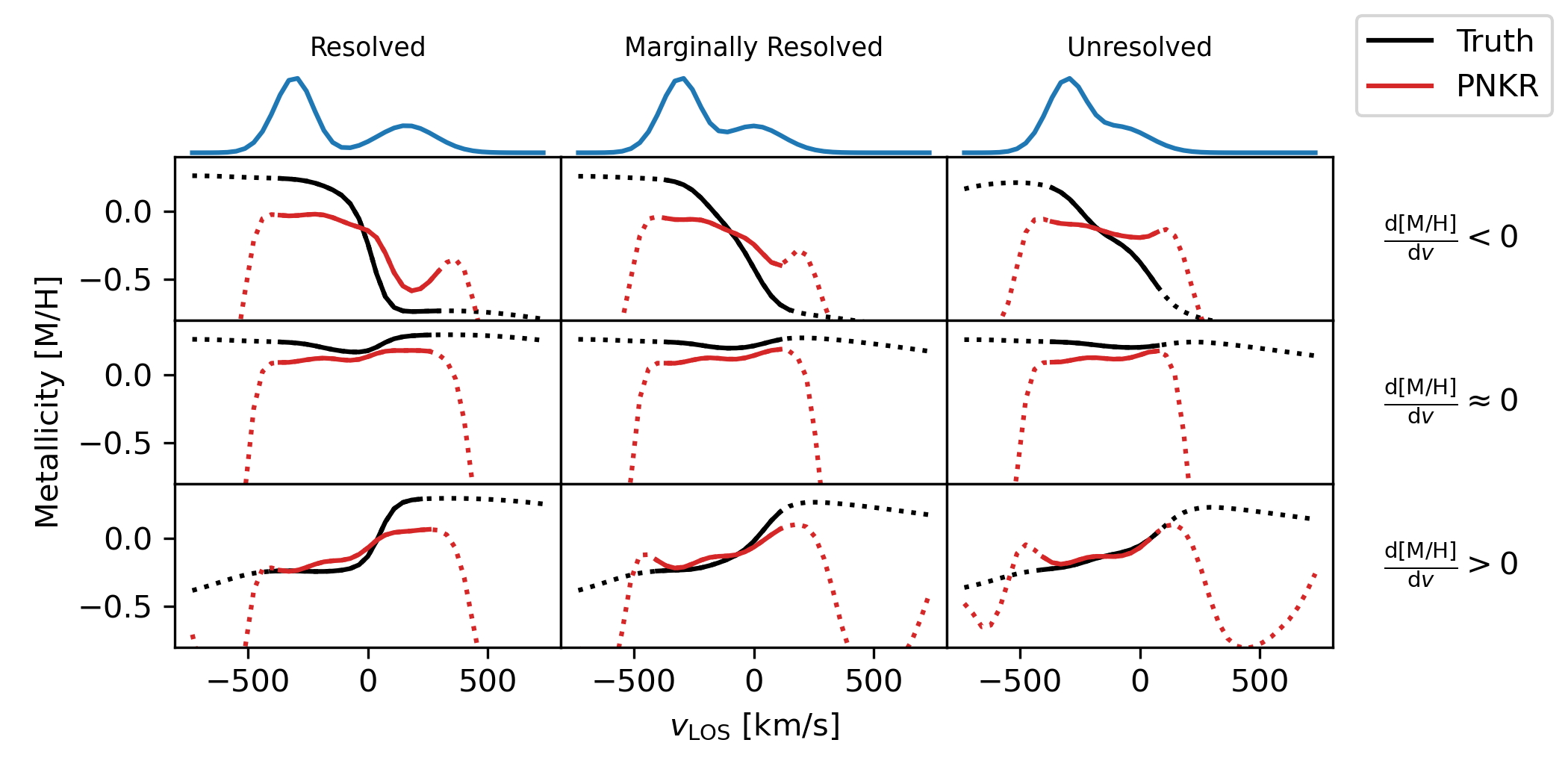}
  \caption{
  Recoveries of spatially-integrated metallicity-velocity relations for 9 different ground-truth models.  
  The models vary in the separation of the two components in velocity space (illustrated by the LOSVDs above each column which correspond to spaxel 7 in Figure~\ref{fig:metal_map}), and the metallicity difference between the two components (in the top/middle/bottom row, metallicity is a decreasing/flat/rising function of velocity).
  In general, the recoveries (red) successfully infer whether the positive-velocity component has metallicity lower than, similar to, or higher than the negative-velocity component.
  There is one exception (top right panel), where a flat metallicity-velocity relation is inferred while the truth is decreasing.
  All nine recoveries are performed at SNR=200.
  Dotted lines show velocities with a negligible contribution, where metallicity is unconstrained.
  }
  \label{fig:metal_vary_deltaZV}
\end{figure*}

Figure~\ref{fig:metal_vary_deltaZV} shows PNKR recoveries of spatially-integrated metallicity-velocity relations for nine different ground truth models.
The models vary in the velocity separation (columns) and metallicity difference (rows) between the two components.
For all nine models, we generate SNR=200 mock data using \texttt{popkinmocks} \citep{Jethwa2023}, and run PNKR-1D recoveries with settings as described in Appendix~\ref{apdx:imp_pnkr1d}.
Examining Figure~\ref{fig:metal_vary_deltaZV}, we see that, although absolute metallicities are often discrepant from the true values, in 8/9 cases, PNKR can successfully infer whether the metallicity-velocity relations are decreasing, flat or rising.
The sole exception is shown in the top right panel, where a flat relation is inferred while the truth is decreasing.
This exception occurs where the two components overlap in velocity, presenting the most challenging recovery problem.
The overall conclusion, however, is that PNKR generally can infer the qualitative behaviour of metallicity-velocity relations with SNR=200 data.

\section{Discussion}
\label{sec:discuss}

In this work, we have tested two algorithms (BLOSVD and PNKR) for non-parametric reconstructions of stellar population/kinematic distributions from IFS data.
A key novelty of both algorithms is that, in addition to modes for 1D spectral fitting (BLOSVD-1D and PNKR-1D), they provide novel functionality to fit multiple spectra simultaneously accounting for spatial correlations (BLOSVD-2D and PNKR-3D).
In this section, we discuss our results and their implications for detecting and characterising stellar structures, for BLOSVD (Section~\ref{ssec:disc_blosvd}) then PNKR (Section~\ref{ssec:disc_PNKR}), before discussing the limitations of this work (Section~\ref{ssec:disc_limit}).

\subsection{BLOSVD}
\label{ssec:disc_blosvd}

Our tests with BLOSVD demonstrate the benefits of accounting for spatial correlations for reconstructing stellar kinematics.
Our LOSVD reconstructions with BLOSVD-2D are both more accurate (Figure~\ref{fig:3d_summary}) and more precise (Figure~\ref{fig:b3d_uncertainties}) than their 1D equivalents.
The improvements in accuracy arise from a reduction in spike artifacts at high SNR and more subtle improvements to the LOSVD shape at lower SNR (Figure~\ref{fig:b3d}).
Regarding precision, we have shown that BLOSVD-2D significantly reduces LOSVD uncertainty at all SNR values tested, and this happens preferentially in the LOSVD body compared to the wings (Figure~\ref{fig:b3d_uncertainties}).
These improvements may enable the detection of weak stellar components which would attain relatively low statistical significance from 1D spectral fits, hence we advocate for the use of some spatial regularisation when modelling IFS data.

While our implementation of spatial regularisation in BLOSVD-2D has demonstrated significant improvements there is scope for further enhancements.
Spatial regularisation has been much more effective in smoothing the dominant ($v_\mathrm{LOS}<0$) component than the sub-dominant ($v_\mathrm{LOS}>0$) component (Figure~\ref{fig:channel_maps}).
We observe that this is related to the $\sigma_\mathrm{CAR}$ hyper-parameter of the CAR prior, in the sense that velocity channels with density $>\sigma_\mathrm{CAR}$ are most effectively smoothed.
This observation motivates a modification to our current implementation, which assigns all velocity bins the same $\sigma_\mathrm{CAR}$, to a more flexible implementation with a velocity-dependent $\sigma_\mathrm{CAR}$.
We caution, however, that $\sigma_\mathrm{CAR}$ strongly effects the stability of MCMC sampling (see Appendix~\ref{apdx:imp_bl3d}), so added flexibility may require additional tuning of the inference algorithm.
Additionally, our current 2D experiment, which is restricted to fitting columns of spaxels, should be extended to a full 3D implementation.
This is a straightforward extension to our current implementation.

In this work, we have shown BLOSVD reconstructions for kinematics but not for stellar-populations.
As described in Section~\ref{ssec:meth_blosvd}, this is because BLOSVD uses PCA to replace several hundred SSP templates with a handful of PCA templates.
BLOSVD treats inferred PCA-weights as nuisance parameters, foregoing all information on stellar-populations, however we note that this is not a fundamental limitation to this approach.
\citet{Parzer23} show how to de-project PCA-weights back to SSP-weights, which can be used to reconstruct physical quantities such as mean ages, metallicities and star formation histories in a post-processing step.
In future work, we will investigate stellar-population properties derived in this way.

Despite it's successes, we have found BLOSVD achieves less accurate LOSVD reconstructions than PNKR for SNR>40, in either its 1D or 2D varieties (Figure~\ref{fig:1d_summary}).
We attribute this result to BLOSVD's ansatz that, locally, stellar-populations and kinematics are independent i.e. equation~\eqref{eqn:simplifying_assumption}.
The assumption introduces a bias to BLOSVD recoveries which can be seen most clearly in the left panel of of Figure~\ref{fig:b3d}: in all spaxels, the amplitude of the $v_\mathrm{LOS}<0$ component is slightly over-predicted, while the $v_\mathrm{LOS}>0$ component is slightly under-predicted.
This trade-off arises since BLOSVD does not have the freedom to assign different stellar-populations to the two components.
PNKR imposes no such assumption, and our experiments with PNKR support the claim that this difference contributes to the relative underperformance of BLOSVD at high SNR.

\subsection{PNKR}
\label{ssec:disc_PNKR}

We have shown that PNKR is able to infer qualitative behaviour of metallicity-velocity relations.
We demonstrate that PNKR can make these measurements locally (Figure~\ref{fig:metal_map}) and, by considering spatially-integrated measurements, show this requires high SNR (Figures \ref{fig:metal_snr}) and is robust against a variety of different ground truth models (Figure~\ref{fig:metal_vary_deltaZV}).
These results provide an explanation as to why PNKR achieves more accurate LOSVD reconstructions than BLOSVD at high SNR, and demonstrate that it is possible to infer dependencies between stellar populations and kinematics in a non-parametric way.

To our knowledge, the only previous work to demonstrate \emph{non-parametric}\footnote{As opposed to \emph{parametric} spectral decompositions \citep[e.g.][]{Coccato11}.} recoveries of joint population-kinematic distributions is \citet{steckmap}.
Figures 5 and 6 of that work illustrate non-parametric recoveries of age-velocity distributions, successfully inferring distinct LOSVDs for components with ages 30 Myr and 3 Gyr.
We extend this result by demonstrating that even given two relatively old components, both with ages > 1 Gyr, PNKR can simultaneously disentangle their kinematics and metallicities.
Measurements such as these will be particularly useful for extragalactic archaeology \citep{ExtragalacticArchaeology}, since by disentangling the properties of old stellar populations, we can learn about a galaxy's evolution over Gyr timescales.

While PNKR is able to well-recover the qualitative behaviour of metallicity-velocity relations at SNR=200, our recovered metallicities are biased by $\approx$ 0.2 dex (Figure~\ref{fig:metal_snr}).
Validation studies of full-spectrum fitting \citep{Koleva08,Ge18,CidFernandes18,Wang24} typically recover mean metallicities with higher accuracy e.g. \citet{Woo24} report a metallicity bias of only 0.07 dex when using pPXF \citep{ppxf04,ppxf17}.
Compared to most validation studies, we model a relatively complex case of a counter-rotating galaxy, however in additional tests with only a single component, the bias did not decrease significantly.
Instead, we believe this bias may be due to PNKR's iterative regularisation scheme.
Just as this caused oversmoothing of LOSVDs (right panel of Figure~\ref{fig:pnkr_vs_blosvd}), PNKR is likely oversmoothing age-metallicity distributions, leading to mean metallicity and ages biased toward to central values of the SSP grid.
While we only observed LOSVD oversmoothing at low SNR, this may be affecting stellar-population distributions even at high SNR due to the smaller number of age/metallicity bins compared to velocity bins in the problem discretisation.
We have identified PNKR's wavelength deactivation scheme as a possible cause of oversmoothing (see the pink line in Figure~\ref{fig:1d_summary}), hence modifications to this scheme (e.g. via wavelength importance sampling) may remedy the issue.
Alternatively, using SSP grids with finer sampling and incorporating prior information on age-metallicity relations may also prove helpful.

We found that accounting for spatial correlations with PNKR-3D did not lead to any improvements compared to PNKR-1D.
This conclusion was valid for LOSVD reconstructions (Figure~\ref{fig:3d_summary}) and metallicity-velocity relations.
The reason for this is unclear.
In \citetalias{Hinterer23}, we proposed two different approaches to incorporate spatial regularisation in PNKR.
In the first approach, the unknown density $f$ is built from \emph{hat-shaped} basis functions which span across adjacent spaxels.
This approach led to some qualitative improvements in LOSVD recoveries (see Figure 5.6 of \citetalias{Hinterer23}) but was prohibitively slow to run.
For this work, we therefore adopted the alternative approach based on spatial smoothing filters described in Section~\ref{ssec:pnkr} and named the \emph{Reduced PNKR Method} in \citetalias{Hinterer23}.
Theoretical arguments suggest that the two approaches are equivalent \citep{Hubmer_Sherina_Ramlau_2023} however we find that the second approach provides little benefit over 1D reconstructions.

\subsection{Limitations}
\label{ssec:disc_limit}

In this work we test the specific scenario of a galaxy with a large-scale stellar counter-rotation.
This scenario was chosen as it is complex enough to illustrate key features of our recovery algorithms, however it is far from typical.
For example, in a systematic search of MaNGA galaxies, \citet{Bevacqua22} found just 64/4000 showed evidence for stellar counter-rotation.
Nevertheless, with the improved spectral resolution of future IFS instruments \citep[e.g.][]{thatte16_harmoni,bluemuse,hector24}, we will be able to disentangle far more typical varieties of stellar kinematic substructures, such as co-rotating disks, bars and bulges.
The benefits of the recovery algorithms shown in this work should generalise to these other situations, however dedicated tests will be useful to understand the limitations in different contexts.

The mock data we have used in this work is idealised.
We have omitted instrumental effects including the PSF, LSF and uncertain flux calibrations, as well as physical effects such as dust-attenuation, nebular emission and sky residuals.
While this suffices for the theoretical investigation we have presented here, progress on real data will require us to deal with these additional effects.
Several existing tools are available to create more realistic mock data \citep[e.g.][]{SIMSPIN,RealSimIFS,PRISMA}.
Both BLOSVD and PNKR can be extended to handle more complex forward models.
Since the PNKR algorithm is tailored to the mathematical structure of the model, modifications to the forward model may require alterations to the inference algorithm.
BLOSVD on the other hand is built in the framework of a probabilistic programming language, where the tasks of model specification and inference are decoupled.
This means additional effects can be added during the modelling process without the need to modify the inference algorithm.

\section{Conclusions}
\label{sec:concs}

We have compared two algorithms for the reconstruction of stellar kinematics and populations from IFS data: PNKR and BLOSVD.
In addition to modes for 1D spectral fitting, both algorithms have extensions which simultaneously fit several spectra accounting for spatial correlations across spaxels.
These are the first implementations of 3D modelling for full spectrum fitting for stellar recoveries.
We evaluate the algorithms using idealised mock data at varying SNR levels.

Accounting for spatial correlations led to significant improvements in the accuracy and precision of LOSVD recoveries with BLOSVD (Section~\ref{ssec:23d_results}).
These benefits may enable the detection of weak stellar kinematic substructures which would evade detection from 1D spectral fits.
We therefore advocate for further development and usage of 3D modelling tools for stellar recoveries.
While we saw significant benefits in BLOSVD, 3D modelling with PNKR did not lead to any noticeable improvements over 1D fits.
We conclude that the implementation of 3D modelling within PNKR, based on smoothing the solution between iterations of the optimisation algorithm, is less effective than that in BLOSVD, which couples adjacent spaxels together using a hierarchical Bayesian prior to encode spatial regularities.
Though our implementation of spatial regularisation in PNKR proved ineffective, it still achieved smaller LOSVD reconstruction errors than BLOSVD for SNR>40 (Figure ~\ref{fig:3d_summary}).
This result can partly be explained by the other key novelty of the PNKR algorithm.

PNKR has the freedom to model the full joint density in stellar populations and kinematics, forgoing the common ansatz that these are locally independent.
With sufficiently high SNR, we find that PNKR makes use of this freedom to successfully recover the qualitative behaviour of metallicity-velocity relations (Section~\ref{ssec:popkin_pnkr}).
Though the qualitative behaviour is well-recovered, absolute metallicities are recovered with a significant bias, which we speculate is caused by PNKR's iterative regularisation scheme.
Nevertheless, this is a promising result which suggests that we may be able to disentangle the population properties of stellar substructures in a wider range of contexts than has been attempted previously.

This work has demonstrated two promising strategies to better detect and characterise stellar structures in galaxies: 3D modelling and joint population-kinematic modelling.
Currently, we have illustrated the benefits of these strategies with two separate recovery algorithms.
In future work, we will integrate both features into a unified algorithm and apply this to existing IFS datasets.

\begin{acknowledgements}
This research was funded in part by the Austrian Science Fund (FWF) SFB 10.55776/F68 ``Tomography Across the Scales'', project F6805-N36 (Tomography in Astronomy) and F6811-N36 (Advancing Extragalactic Archaeology through Novel Inversion Techniques). For open access purposes, the authors have applied a CC BY public copyright license to any author-accepted manuscript version arising from this submission. The computational results have been achieved using the Austrian Scientific Computing (ASC) infrastructure. The authors thank Dr. Fabian Hinterer for his contributions.

\end{acknowledgements}

\bibliographystyle{aa}
\bibliography{pnkr4dummies}

\begin{appendix}

\section{Additional Details}
\label{apdx:implementation}

This appendix contains implementation details and comments on the convergence of the recoveries presented in Section~\ref{sec:results}.

\subsection{PNKR-1D}
\label{apdx:imp_pnkr1d}

For PNKR-1D recoveries in Section~\ref{ssec:1d_results}, we use step-size 100 and tolerance of 1.3 to determine wavelength deactivation i.e. if a wavelength is fit to within 30\% accuracy it is deactivated.
All PNKR-1D fits reach zero active equations before the algorithm terminates.
The number of iterations required to reach this point increased with SNR, from 74 iterations for SNR=20 to 560 for SNR=200.
Correspondingly, computational run times increased with SNR, from 1 hour at SNR=20 to 15 hours for SNR=200.
In attempts to to fix LOSVD oversmoothing for SNR=20, we additionally ran recoveries with smaller step-sizes (50, 10) and a less generous tolerance (1.1) but found these led to no improvement.

\subsection{PNKR-3D}
\label{apdx:imp_pnkr3d}

For PNKR-3D recoveries, we use the same setting and tolerance listed in section \ref{apdx:imp_pnkr1d}.
For the parameters of the smoothing kernel, we use a (5$\times$5) grid over shape parameter $\alpha$ and scale-length $\beta$.
We take 5 values of $\alpha$ spaced linearly between 0.2 and 1.0.
For SNR=200, we take 5 values of $\beta$ logarithmically spaced between $10^{-5}$ and $10^{-3}$.
For lower SNR, we inversely scale the $\beta$ values with SNR, so that lower SNR data can benefit from more spatial regularisation.

\begin{figure*}
  \centering
  \includegraphics[width=\textwidth]{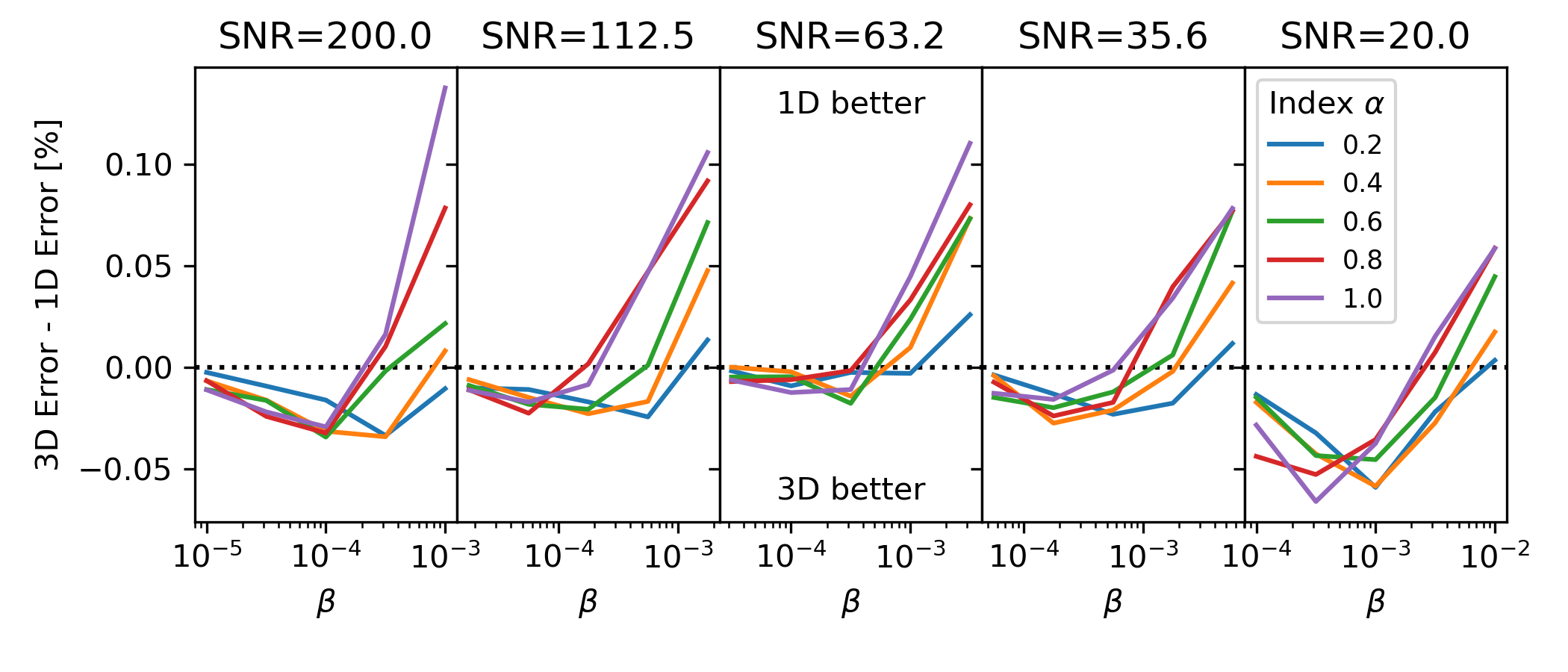}
  \caption{
  Panels show the change in LOSVD recovery error when using PNKR-3D vs PNKR-1D (y-axis) vs. the kernel scale length $\beta$ (x-axes) and power-law index $\alpha$ (colours as shown in legend).
  The 5 panels show this for different SNR levels, decreasing from left to right.
  }
  \label{fig:pnkr_error_3d_vs_1d}
\end{figure*}

Figure~\ref{fig:pnkr_error_3d_vs_1d} shows the change in LOSVD recovery error for PNKR-3D vs PNKR-1D.
For all SNR and for all choices of $\alpha$, we see that the difference in recovery error shows a U-shape: for small $\beta$, the error difference approaches 0, while for large $\beta$, PNKR-3D becomes oversmoothed and it's error exceeds PNKR-1D.
These U-shape curves reach a minimum for some intermediate $\beta$.
This gives us confidence that the range of smoothing parameters we have explored is sufficient i.e. our conclusion that PNKR-3D provides little benefit is not limited by this range.

Regarding convergence, in contrast to PNKR-1D fits, where we always reached zero active equations, PNKR-3D fits instead stopped at a plateau defined by no decrease in the number of active equations over the last 100 iterations.
The number of iterations required to reach the plateau increases with SNR, from 200 at SNR=20 to 350 for SNR=200, with small variations (approx 10) depending on the smoothing kernel parameters.
For SNR=200 we ran additional tests with a plateau defined by no decrease in the number of active equations over the last 300 iterations, but found no change.

\subsection{BLOSVD-1D}
\label{apdx:imp_bl1d}

For BLOSVD-1D, we use 15 PCA templates to represent the SSP grid (which accounts for $>99\%$ of the variance of the full grid), and run the NUTS MCMC sampler for 1000 warmup steps followed by 2000 sampling steps.
For all SNR, BLOSVD-1D fits successfully achieve the converge criteria in Section \ref{ssec:meth_blosvd}).
Run times were 30 minutes at high SNR and 3 minutes at low SNR.
These times are much smaller than those for PNKR-1D since we can parallelise over the spaxels.

\subsection{BLOSVD-2D}
\label{apdx:imp_bl3d}

For BLOSVD-2D we use again use 15 PCA templates to represent the SSP grid, and again run the NUTS MCMC sampler for 1000 warmup steps followed by 2000 sampling steps.
In this case, however, when running the NUTS sampler, we found that using including all spaxels led to a prohibitively slow warmup stage.
We therefore limited ourselves to using only a fifth of the spaxels (i.e. 6 out of 30 spaxels per column) for the warmup stage, took the resulting post-warmup settings and used these to initiate the sampling stage including all spaxels.
The initial positions of the MCMC chains are  posterior samples from 1D fits from each spaxel.

The CAR hyperparameters discussed in Section \ref{ssec:meth_blosvd} are set as follows.
For the correlation parameter $0<\rho_\mathrm{CAR}<1$ we take a hyperprior strongly peaked at $\rho_\mathrm{CAR}=1$ to promote smoothness.
Concretely, we take $p(\rho_\mathrm{CAR}) = \mathrm{Beta}(\rho_\mathrm{CAR};4,1)$ which has a single mode at $\rho=1$ and a median at $\rho_\mathrm{CAR}=0.84$.
We encountered poor convergence diagnostics when we used a hyperprior on the second CAR hyperparameter $\sigma_\mathrm{CAR}$.
Specifically, when using a uniform prior on $\log \sigma_\mathrm{CAR}$, we encountered many divergent transitions when MCMC chains approached $\sigma_\mathrm{CAR}=0$.
When truncating the prior below some value $\sigma_\mathrm{CAR}>0$, the resulting samples invariably bunched up tightly against the truncation limit.
To have more control over these experiments, we therefore abandoned the use of a hyperprior on $\sigma_\mathrm{CAR}$ and instead tried two fixed values $\sigma_\mathrm{CAR} \in [0.001, 0.03]$.

The convergence of the resulting BLOSVD-2D depended on the choice of $\sigma_\mathrm{CAR}$.
All fits converged successfully for $\sigma_\mathrm{CAR}=0.03$.
For $\sigma_\mathrm{CAR}=0.001$ many fits failed, exhibiting poor mixing and small effective sample sizes.
Out 30 columns of spaxels, we encountered [30,6,5,1,0] failures for our five SNR bins increasing from 20 to 200 i.e. all columns failed at SNR=20 while all succeeded for SNR=200.
For the results shown in Section \ref{sec:results} we required all columns to converge successfully, therefore for SNR<200 we show only results with $\sigma_\mathrm{CAR} = 0.03$.

\section{Quality of Fit}
\label{apdx:qof}

\begin{figure*}
  \centering
  \includegraphics[width=\textwidth]{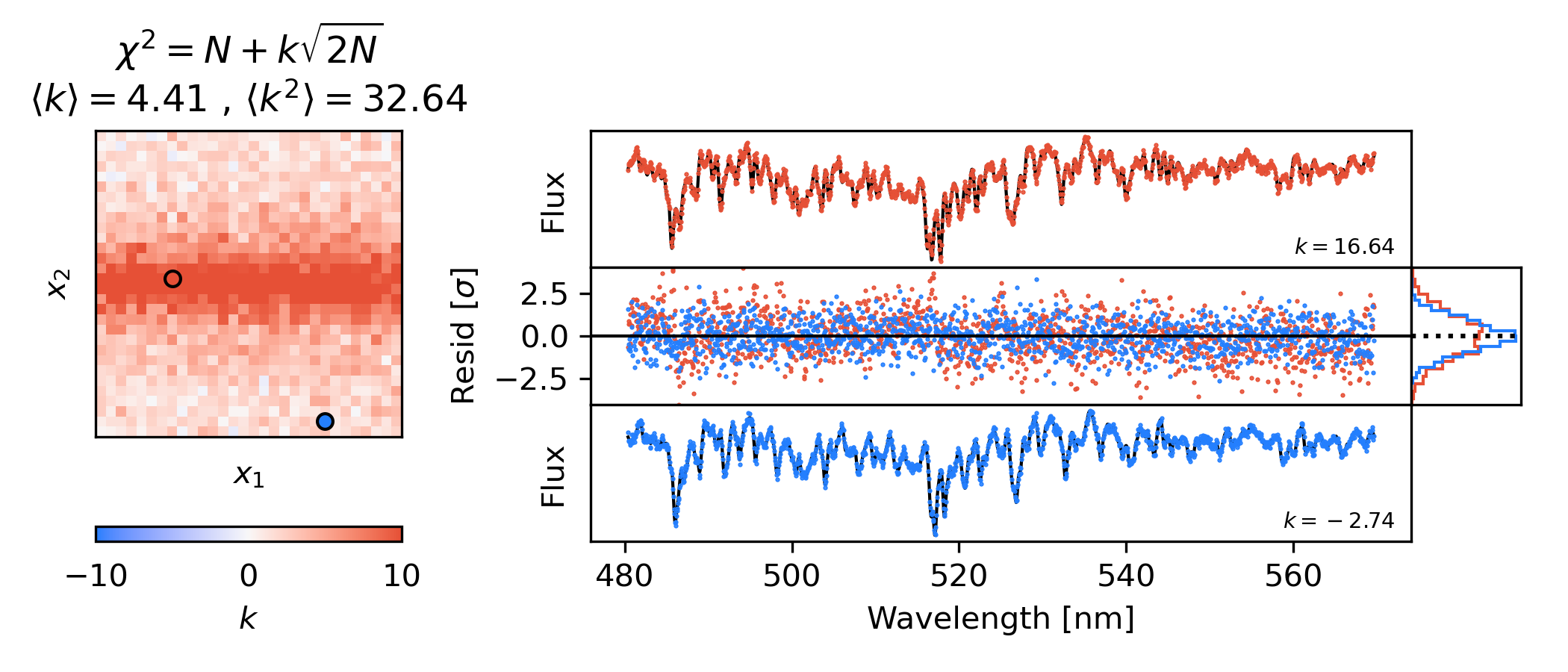}
  \caption{
  Quality of fit for PNKR-1D at SNR=200. Figure described in Section \ref{apdx:qof}.
  }
  \label{fig:qof_pnkr200}
\end{figure*}

\begin{figure*}
  \centering
  \includegraphics[width=\textwidth]{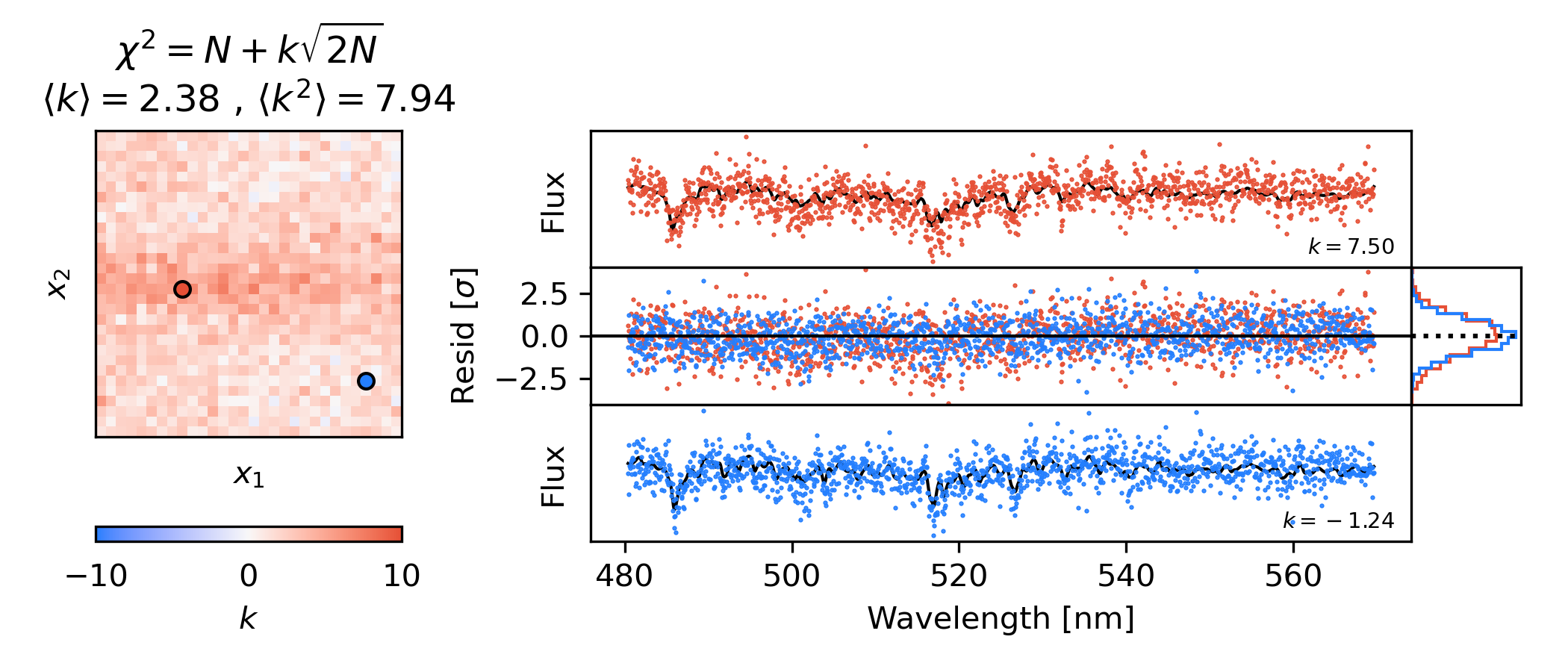}
  \caption{
  Quality of fit for PNKR-1D at SNR=20. Figure described in Section \ref{apdx:qof}.
  }
  \label{fig:qof_pnkr20}
\end{figure*}

\begin{figure*}
  \centering
  \includegraphics[width=\textwidth]{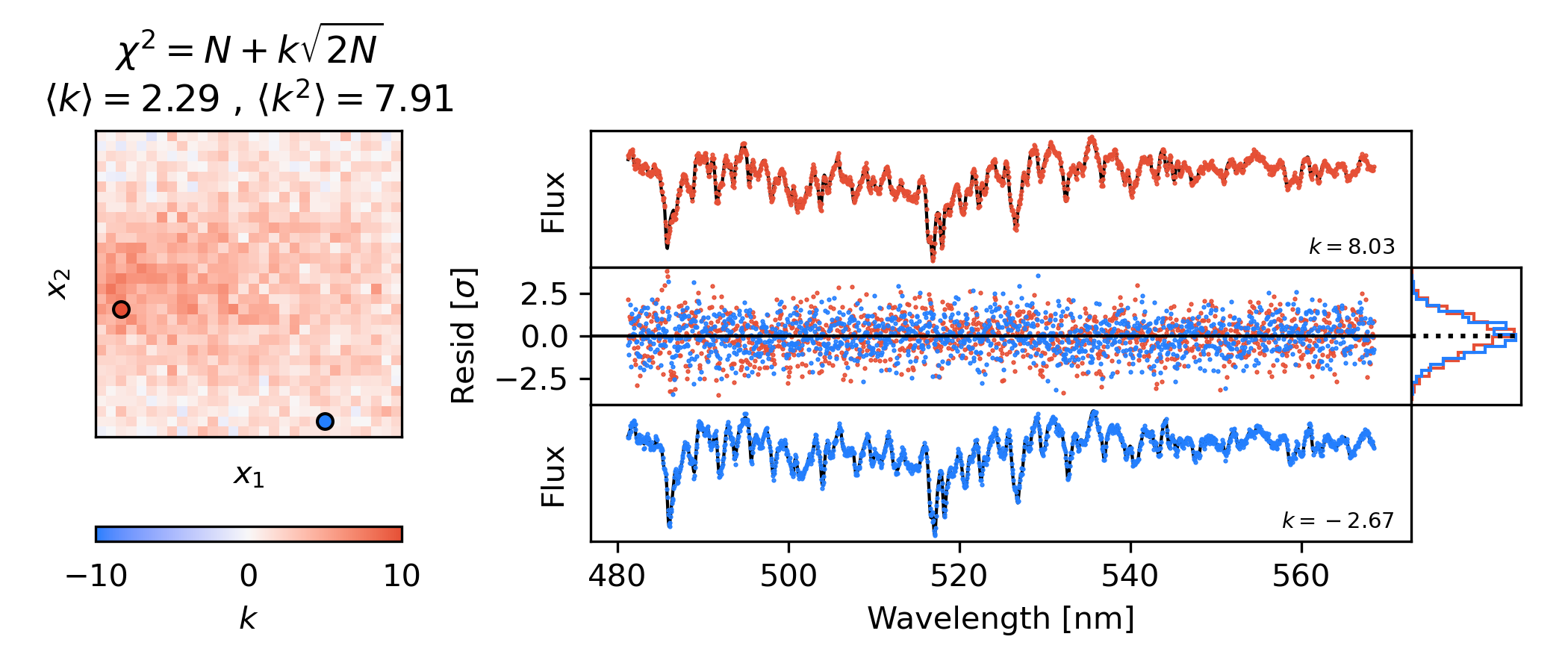}
  \caption{
  Quality of fit for BLOSVD-1D at SNR=200. Figure described in Section \ref{apdx:qof}.
  }
  \label{fig:qof_blosvd200}
\end{figure*}

\begin{figure*}
  \centering
  \includegraphics[width=\textwidth]{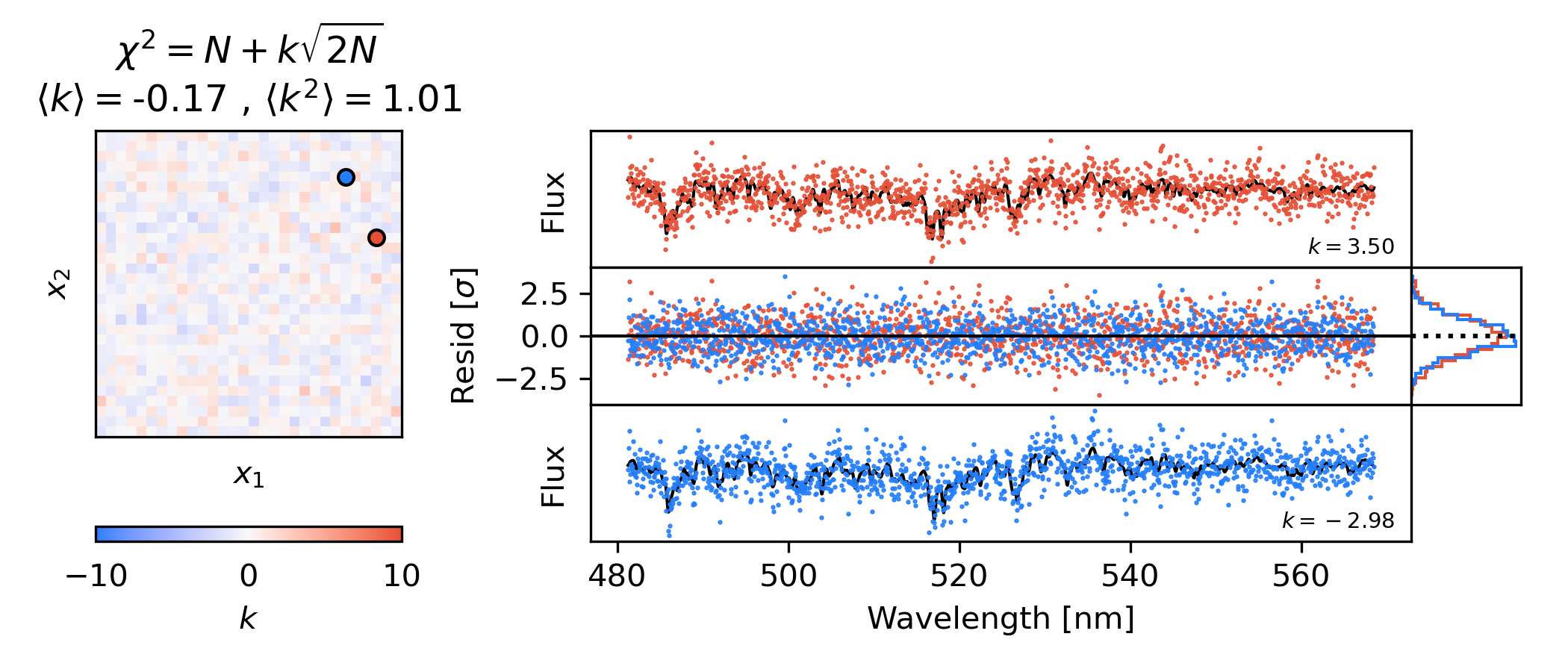}
  \caption{
  Quality of fit for BLOSVD-1D at SNR=20. Figure described in Section \ref{apdx:qof}.
  }
  \label{fig:qof_blosvd20}
\end{figure*}

\begin{figure*}
  \centering
  \includegraphics[width=\textwidth]{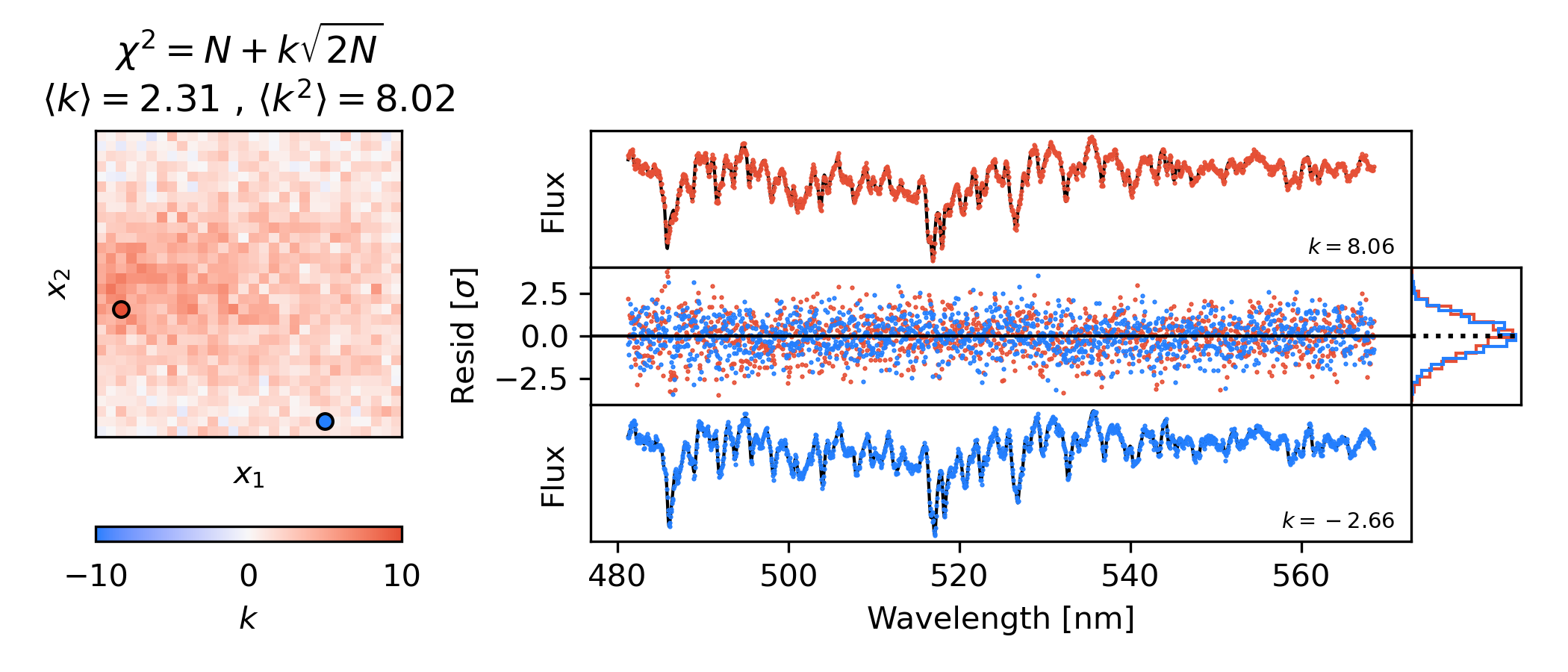}
  \caption{
  Quality of fit for BLOSVD-2D at SNR=200. Figure described in Section \ref{apdx:qof}.
  }
  \label{fig:qof_blosvd2d200}
\end{figure*}

\begin{figure*}
  \centering
  \includegraphics[width=\textwidth]{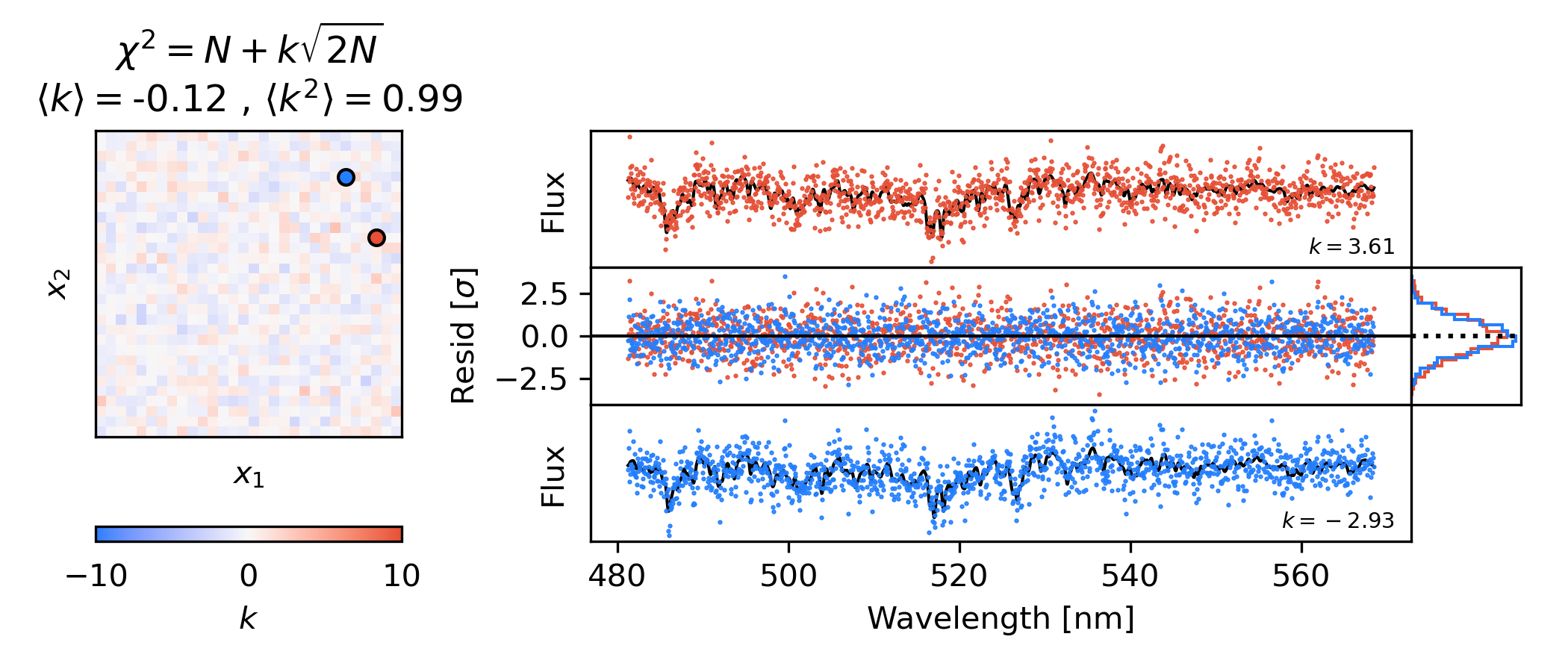}
  \caption{
  Quality of fit for BLOSVD-2D at SNR=20. Figure described in Section \ref{apdx:qof}.
  }
  \label{fig:qof_blosvd2d20}
\end{figure*}

Figures \ref{fig:qof_pnkr200} to \ref{fig:qof_blosvd2d20} visualise the quality of fit for (1) PNKR-1D at SNR=200, (2) PNKR-1D at SNR=20, (3) BLOSVD-1D at SNR=200, (4) BLOSVD-1D at SNR=20, (5) BLOSVD-2D at SNR=200, (6) BLOSVD-2D at SNR=20.
In each case, the right panels show two spectra selected at the best fit (blue, bottom row) and worst fit (red, top row) spaxels, with the residuals normalised by $\sigma$ in the middle row.
The images on the left show $\chi^2$ maps between observed and reconstructed datacubes, displayed as the difference $k$ from the number of wavelength bins $N$ in units of $\sqrt{2N}$.
In these units, a \emph{perfect} fit would centre at $\left<k\right>=0$ with a variance $\left<k^2\right>=1$ and display no spatial structure.
Both BLOSVD-1D and 2D approach this ideal scenario for SNR=20 (Figures \ref{fig:qof_blosvd20} and \ref{fig:qof_blosvd2d20}) however at SNR=200 the $\chi^2$ maps show some spatial structure.
PNKR-1D reconstructions (Figures \ref{fig:qof_pnkr200} and \ref{fig:qof_pnkr20})) lie further from the observations than BLOSVD-1D with significantly more spatial structure.
This is to be expected due to PNKR's implicit regularisation scheme, which trades fidelity of the data reconstruction for accuracy in the reconstruction of the unknown density.
Nevertheless, in all cases, the example spectra in the right panels show that the reconstructed spectra (black lines) do tend to trace the data (coloured dots), giving us confidence that all algorithms tested are functioning as expected.

\section{Definitions}

This appendix defines several quantities referenced throughout this work.

\subsection{Light Weighted LOSVDs}
\label{apdx:lw_losvds}

The distribution $f(\xv,v,z,t)$ defined in equation~\eqref{eqn:forward model} is mass-weighted, since SSPs $S(\lambda,z,t)$ are defined as spectra per unit-mass of a stellar population.
The equivalent light-weighted distribution $f^\mathcal{L}$ is achieved by scaling by the integrated flux of the SSPs i.e.
\begin{equation}
    f^\mathcal{L}(\xv,v,z,t) \propto f(\xv,v,z,t) \int S(\lambda,z,t) \;\mathrm{d}\lambda \;.
    \label{eqn:light_weighted_density}
\end{equation}
The light-weighted LOSVD at position $\xv$ is then given by 
\begin{equation}
    f^\mathcal{L}(v|\xv) 
        = 
        \frac{
        \int f^\mathcal{L}(\xv,v,z,t) \;\mathrm{d}z \;\mathrm{d}t
        }{
        \int f^\mathcal{L}(\xv,v,z,t) \;\mathrm{d}v \;\mathrm{d}z \;\mathrm{d}t
        }.
    \label{eqn:light_weighted_losvds}
\end{equation}
Note that BLOSVD adopts a simplifying assumption - equation~\eqref{eqn:simplifying_assumption} - which imposes that, locally, stellar populations and kinematics are independent, and therefore light-weighted or mass-weighted LOSVDs are, by definition, identical.

\subsection{LOSVD Recovery Error}
\label{apdx:losvd_rec_err}

Given the true light-weighted LOSVDs $f^\mathcal{L}_\mathrm{true}(v|\xv)$ and a recovery $f^\mathcal{L}_\mathrm{rec}(v|\xv)$, we define the error between them as
\begin{equation}
    \mathrm{Error} = \frac{1}{N_\mathrm{spax}} \int \int \left| \frac{f^\mathcal{L}_\mathrm{true}(v|\xv) - f^\mathcal{L}_\mathrm{rec}(v|\xv)}{f^\mathcal{L}_\mathrm{true}(v|\xv)} \right| f^\mathcal{L}_\mathrm{true}(v|\xv) \;\mathrm{d}\xv \;\mathrm{d}v \;.
    \label{eqn:losvd_recov_error}
\end{equation}
This can be interpreted as the fractional error, weighted by the true LOSVD, averaged over all spaxels and velocity bins.
Note that - by definition in equation~\eqref{eqn:light_weighted_losvds} - the true LOSVDs are probability density functions over $v$ at each $\xv$, and hence they are appropriately normalised to serve as weighting factors in this definition.
In practice, we calculate this error ignoring the two cancelling factors of $f^\mathcal{L}_\mathrm{true}(v|\xv)$, to avoids numerical issues, and approximate the integrals as discrete sums over spaxels and velocity bins.

\subsection{Local Velocity-Metallicity Relations}
\label{apdx:local_vel_met}

The local velocity-metallicity relations shown in Section \ref{ssec:popkin_pnkr} are light-weighted mean metallicities conditional on position and velocity - i.e. $\mathbb{E}(z|v,\xv)$ - evaluated for some fixed choice of position $\xv$.
To define $\mathbb{E}(z|v,\xv)$ we start from the light-weighted joint density $f^\mathcal{L}(\xv,v,z,t)$ defined in equation~\eqref{eqn:light_weighted_density}.
The light-weighted metallicity distribution conditional on velocity and position is then given by
\begin{equation}
    f^\mathcal{L}(z|v,\xv)
        = 
        \frac{
        \int f^\mathcal{L}(\xv,v,z,t) \;\mathrm{d}t
        }{
        \int f^\mathcal{L}(\xv,v,z,t) \;\mathrm{d}z \;\mathrm{d}t
        }.
    \label{eqn:lw_metallicity_vs_vx}
\end{equation}
Finally, the light-weighted mean metallicity conditional on velocity and position is given by
\begin{equation}
    \mathbb{E}(z|v,\xv) = \int z f^\mathcal{L}(z|v,\xv) \;\mathrm{d}z.
    \label{eqn:mean_z_vs_v_x}
\end{equation}

\subsection{Global Velocity-Metallicity Relation}
\label{apdx:global_vel_met}

The global velocity-metallicity relations shown in Section \ref{ssec:popkin_pnkr} are light-weighted mean metallicities conditional on velocity.
This is calculated in a similar way as the local relations defined in the previous section, with additional integrals over position in the numerator and denominator of equation \eqref{eqn:lw_metallicity_vs_vx}.

\section{Age-Velocity Relations}
\label{apdx:age_vel}

\begin{figure*}
  \centering
  \includegraphics[width=\textwidth]{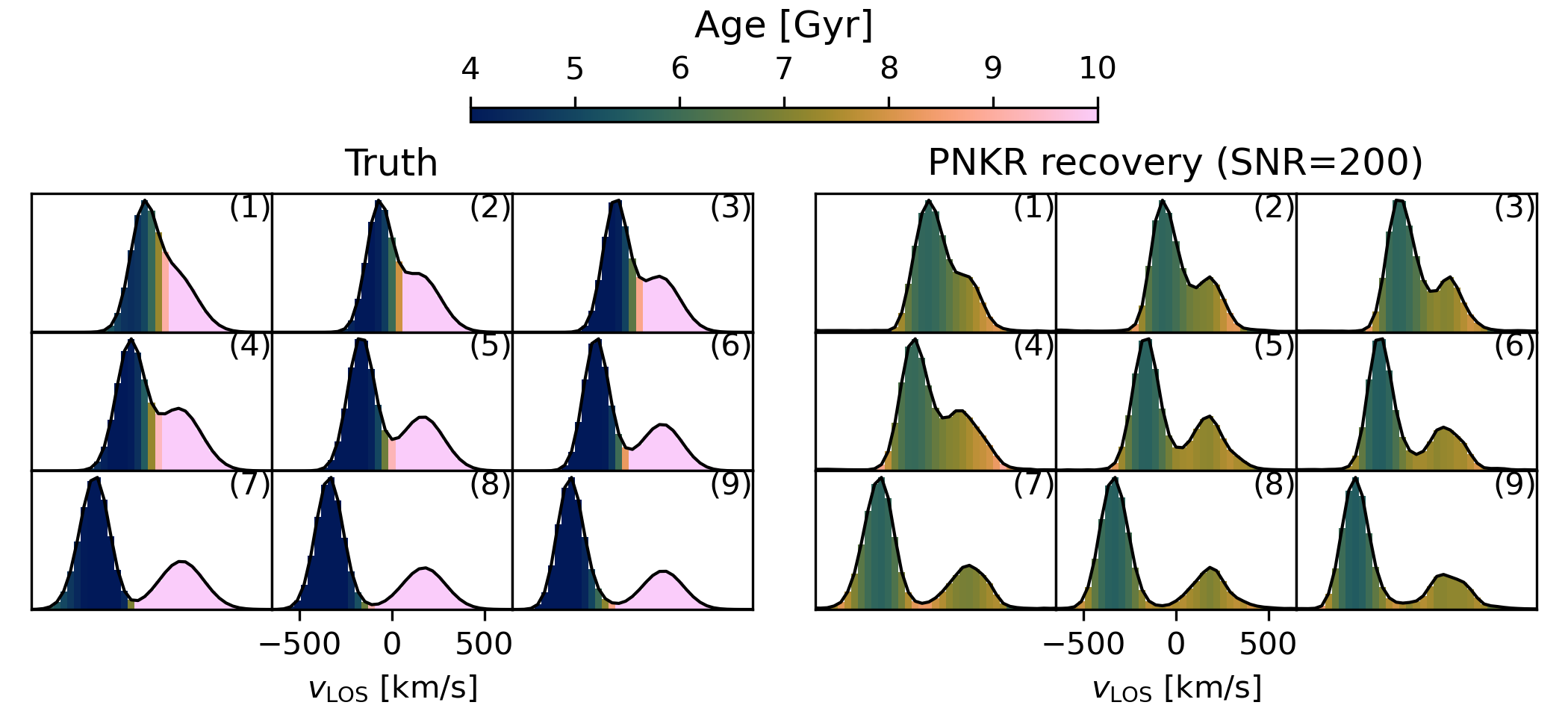}
  \caption{
  Local age-velocity relations recovered with PNKR-1D at SNR=200.
  }
  \label{fig:age_map}
\end{figure*}

\begin{figure}
  \centering
  \includegraphics{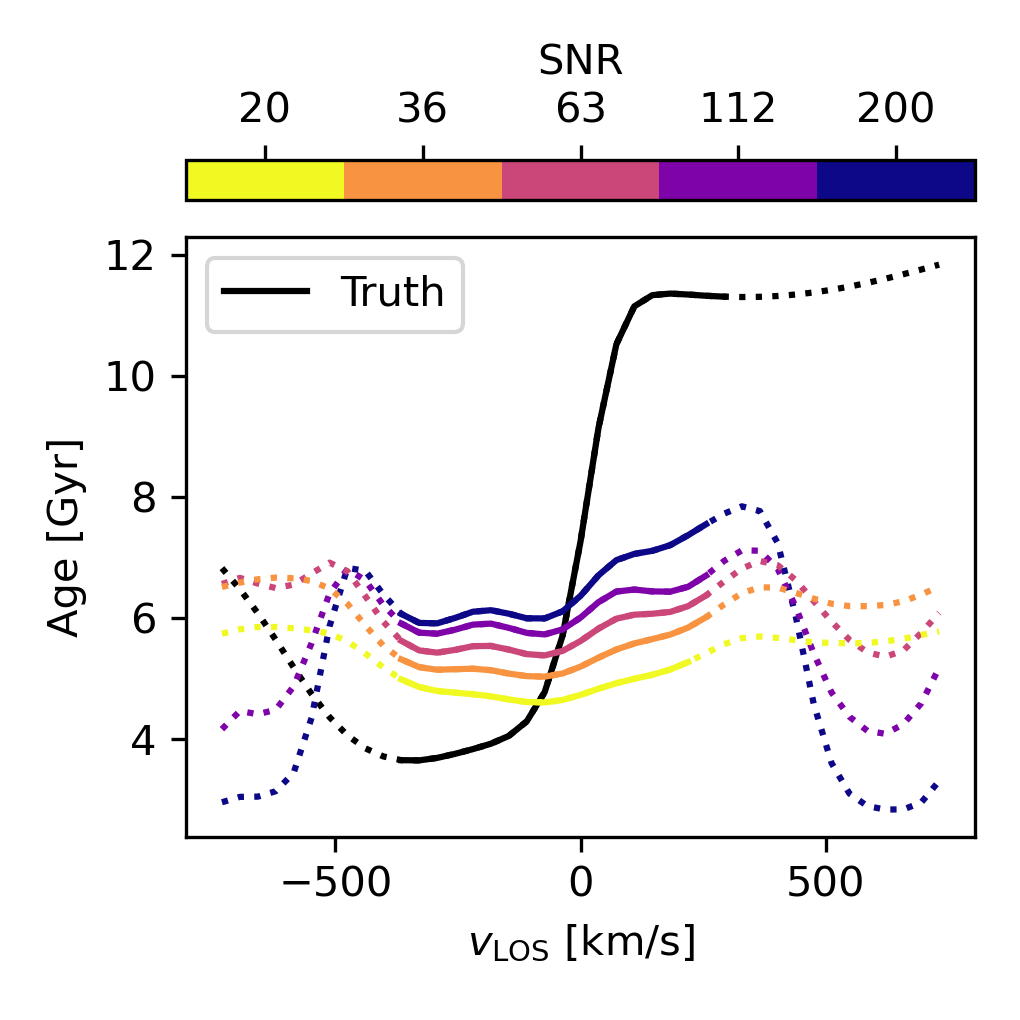}
  \caption{
  Global age-velocity relations recovered with PNKR-1D over a range of SNR.
  }
  \label{fig:age_snr}
\end{figure}

This appendix shows age-velocity relations equivalent to the metallicity-velocity relations shown in Section \ref{ssec:popkin_pnkr}.
Figure \ref{fig:age_map} shows the local relations (equivalent to Figure \ref{fig:metal_map}) while Figure \ref{fig:age_snr} shows global relations at a range of SNR (equivalent to Figure \ref{fig:metal_snr}).
For both cases, the results are qualitatively similar to those seen in in Section \ref{ssec:popkin_pnkr} however the tendency for PNKR to push the recovered ages towards the middle of the age range present in the SSP grid is stronger than for metallicity.
For example, we see this in Figure \ref{fig:age_map}, where PNKR correctly infers that the dominant component is youngest, however the distinction between the ages of the two components is much blurrier than the clear metallicity difference seen in Figure \ref{fig:metal_map}.

\end{appendix}

\end{document}